\documentclass[prb,aps,twocolumn]{revtex4}
\usepackage{latexsym}
\usepackage{graphicx}
\usepackage{subfigure} 
\usepackage{placeins}
\usepackage{amsmath}

\begin{document}

\title{ Coincidence angle-resolved photoemission spectroscopy: Proposal for detection of two-particle correlations }

\author{Yuehua Su} 
\email{suyh@ytu.edu.cn}
\affiliation{ Department of Physics, Yantai University, Yantai 264005, People's Republic of China }

\author{Chao Zhang}
\affiliation{ Department of Physics, Yantai University, Yantai 264005, People's Republic of China }

\begin{abstract}

The angle-resolved photoemission spectroscopy (ARPES) is one powerful experimental technique to study the electronic structure of materials. As many electron materials show unusual many-body correlations, the technique to detect directly these many-body correlations will play important roles in the study of their many-body physics. In this article, we propose a technique to detect directly the two-particle correlations, a coincidence ARPES (cARPES) where two incident photons excite two respective photoelectrons which are detected in coincidence. While the one-photon-absorption and one-photoelectron-emission ARPES provides the single-particle spectrum function, the proposed cARPES with two-photon absorption and two-photoelectron emission is relevant to a two-particle Bethe-Salpeter wave function. Examples of the coincidence detection probability of the cARPES for a free Fermi gas and a Bardeen-Cooper-Schrieffer (BCS) superconducting state are studied in detail. We also propose another two experimental techniques, a coincidence angle-resolved photoemission and inverse-photoemission spectroscopy (cARP/IPES) and a coincidence angle-resolved inverse-photoemission spectroscopy (cARIPES). As all of these proposed coincidence techniques can provide the two-particle frequency Bethe-Salpeter wave functions, they can show the momentum and energy dependent two-particle dynamical physics of the material electrons in the particle-particle or particle-hole channel. Thus, they can be introduced to study the Cooper-pair physics in the superconductor, the itinerant magnetism in the metallic ferromagnet/antiferromagnet, and the particle-hole pair physics in the metallic nematic state. Moreover, as the two-particle Bethe-Salpeter wave functions also involve the inner-pair dynamical physics, these proposed coincidence techniques can be used to study the inner-pair time-retarded physics.    

\end{abstract}


\maketitle

\section{Introduction}\label{sec1}

The most dramatic features of the strongly correlated electron materials, such as the unconventional superconductors of cuprates\citep{PALeeRMP2006}, iron-based superconductors\citep{ChenDaiFeSCs2014,StewartFeSCRMP2011} and heavy fermions\citep{StewartNFLRMP2001,ColemanReview2015}, are the many-body correlations beyond the Landau Fermi liquid physics. These include such as the physics of the Cooper pairs in the superconductor, the itinerant magnetism in the metallic ferromagnet/antiferromagnet, and the particle-hole pair physics in the metallic Pomeranchuk or bond nematic state of the iron-based superconductors\citep{Fradkin,SuLi2015,SuLi2017}. The non-Fermi liquid physics, such as the strange metallic state or the quantum criticality, are ubiquitous in strongly correlated electron materials.\citep{Fradkin,VarmaPhysRep2002,LohneysenRMP2007,SuLu2018}         

Various different experimental techniques have been introduced to study the novel many-body physics in these electron materials. The charge resistivity, the Hall conductivity, and the dynamical optical conductivity show charge current responses. The static magnetic susceptibility, the inelastic neutron scattering, and the nuclear magnetic resonance provide magnetic responses. The ARPES and the scanning tunneling microscope present the electronic single-particle spectrum function and the local density of states, respectively. In all of these experimental techniques in the study of the superconducting Cooper pairs, the itinerant magnetic moments and the nematic charge particle-hole pairs, the inner-pair two-particle correlations of the material electrons can only be inferred indirectly. 

In this article, we will propose a cARPES to detect directly the two-particle correlations. The experimental installation of a cARPES has two photon sources and two photoelectron detectors with an additional coincidence detector. When two photons are incident on a sample material, two electrons can absorb these two photons and can emit outside the sample material as photoelectrons if their energies are high enough to overcome the material work function. The two photoelectrons are then detected in coincidence by the coincidence detector with the coincidence counting probability relevant to a two-particle 
Bethe-Salpeter wave function.

The two-particle Bethe-Salpeter wave function for the cARPES is defined as $\langle \Psi^{s}_{\beta} \arrowvert T_t c_{\mathbf{k}_2 \sigma_2} \left(t_2\right) c_{\mathbf{k}_1 \sigma_1} \left(t_1\right) \arrowvert \Psi^{s}_{\alpha} \rangle$, where $\vert \Psi^{s}_{\alpha} \rangle$ and $\vert \Psi^{s}_{\beta} \rangle$ are the eigenstates of the sample electrons, $c_{\mathbf{k} \sigma}$ is the annihilation operator with momentum $\mathbf{k}$ and spin $\sigma$, and $T_t$ is a time-ordering operator. This Bethe-Salpeter wave function describes the physics of the sample electrons when two electrons are annihilated in time ordering. Therefore, it describes the dynamical physics of the sample electrons with one particle-particle pair (more exactly, hole-hole pair). The cARPES can provide directly the frequency Fourier-transformed Bethe-Salpeter wave function, which shows the momentum and energy resolved particle-particle pair dynamical physics of the sample electrons, including the center-of-mass and inner-pair relative dynamics. Thus, it can be introduced to study the two-particle correlations in the particle-particle channel, such as the Cooper-pair physics in the superconductor. 

We will also propose another two experimental techniques to detect directly the two-particle correlations, a cARP/IPES and a cARIPES. In a cARP/IPES, there are one photon source and one electron source. While an incident photon is absorbed by a sample electron which can emit into vacuum to be a photoelectron, an incident electron with high energy can transit into a low-energy state of the sample material with a photon emitting simultaneously. A coincidence detector then counts the coincidence probability of the photoelectron and the emitting photon, which involves a particle-hole Bethe-Salpeter wave function of the sample electrons, $\langle \Psi^{s}_{\beta} \arrowvert T_t c^{\dag}_{\mathbf{k}_2 \sigma_2} \left(t_2\right) c_{\mathbf{k}_1 \sigma_1} \left(t_1\right) \arrowvert \Psi^{s}_{\alpha} \rangle$. Thus, the cARP/IPES can provide the dynamical physics of the sample electrons with one particle-hole pair. In the spin channel, it can show the information on the itinerant magnetic moments in the metallic ferromagnet/antiferromagnet, and in the charge channel, it can present the information on the particle-hole pairs in the metallic nematic state. In a cARIPES, there are two electrons which are incident on the sample material. They can transit into the low-energy states of the sample electrons with two photons emitting simultaneously. There is a coincidence detector which records the two emitting photons in coincidence with the counting probability being relevant to a two-particle Bethe-Salpeter wave function in the particle-particle channel, $\langle \Psi^{s}_{\beta} \arrowvert T_t c^{\dag}_{\mathbf{k}_2 \sigma_2} \left(t_2\right) c^{\dag}_{\mathbf{k}_1 \sigma_1} \left(t_1\right) \arrowvert \Psi^{s}_{\alpha} \rangle$. As this two-particle Bethe-Salpeter wave function involves mainly the electronic states above the Fermi energy, the cARIPES can show the particle-particle pair dynamical physics of the sample electrons, such as the Cooper pairs in the superconductor, with the electron energies mainly above the Fermi level.   

One special remark is that all of the above three proposed coincidence detection techniques can provide the inner-pair dynamical physics of the sample electrons. Thus, they can be introduced to study the time-retarded dynamics of the two-particle correlations in the particle-particle or particle-hole channel. They may play unusual roles in the study of the dynamical formation of the Cooper pair due to the retarded electron-electron attraction, or the microscopic formation of the itinerant magnetic moment in the metallic ferromagnet/antiferromagnet. 

Our article will be arranged as below. In the following Sec. \ref{sec2}, the theoretical formalism for the cARPES will be established. In Sec. \ref{sec3} the cARPES spectra of a free Fermi gas and a BCS superconducting state will be presented. The theoretical formalisms for the cARP/IPES and cARIPES will be provided in Sec. \ref{sec4}, where the coincidence probability in a contour-time ordering formalism will also be simply discussed. A summary will be presented in Sec. \ref{sec5}.

\section{Theoretical formalism for $\text{c}$ARPES}\label{sec2}

In this section we will establish the theoretical formalism for the cARPES which detects the two-particle correlations in the particle-particle channel. First, we will review the electron-photon interaction in Sec. \ref{sec2.1} and the ARPES in Sec. \ref{sec2.2}. We will then provide the theoretical formalism for the cARPES in Sec. \ref{sec2.3}.

\subsection{Electron-photon interaction}\label{sec2.1}

The lattice model with an external electromagnetic vector potential $\mathbf{A}$ has a kinetic Hamiltonian
\begin{equation}
H(\mathbf{A}) = -\sum_{ij\sigma} t_{ij} e^{i\frac{e}{\hbar} \mathbf{A}_{ij}\cdot \left(\mathbf{R}_j - \mathbf{R}_i\right) } c^{\dag}_{i\sigma} c_{j\sigma} , \label{eqn2.1.1}
\end{equation} 
where the electron charge $q_e = - e$ and the vector potential is defined on-bond $\mathbf{A}_{ij}=\mathbf{A}\left[\frac{1}{2}\left( \mathbf{R}_i + \mathbf{R}_j\right)\right]$. For one single photon mode with $\mathbf{A}_{ij}=\mathbf{A}(\mathbf{q}) e^{i\frac{1}{2}\mathbf{q}\cdot \left(\mathbf{R}_i+\mathbf{R}_j\right)}$, the electron-photon interaction is obtained by a linear-$\mathbf{A}$ expansion of $H(\mathbf{A})$, 
\begin{equation}
V = - \sum_{\mathbf{k}\sigma} \mathbf{v}(\mathbf{k,q}) \cdot\mathbf{A(q)} c^{\dag}_{\mathbf{k+q}\sigma} c_{\mathbf{k}\sigma}  ,  \label{eqn2.1.2}
\end{equation}
where the charged velocity $\mathbf{v}$ is given by
\begin{equation}
\mathbf{v}(\mathbf{k,q}) = \sum_{\mathbf{\delta}} \frac{i e}{\hbar} t_{i,i+\delta} \boldsymbol{\delta} e^{i (\mathbf{k+\frac{q}{2}})\cdot\boldsymbol{\delta}} . \label{eqn2.1.3}
\end{equation}
In the above definitions, $\mathbf{k}$ and $\mathbf{q}$ are momenta and $\sigma$ denotes the electron spin. This electron-photon interaction has only linear-$A$ expansion of $H(\mathbf{A})$, which involves only one-photon emission or absorption in the electron-photon interaction vertex. The quadratic expansion of $H(\mathbf{A})$ with a form as $ \vert \mathbf{A}\vert^2 c^{\dag} c $ involves two-photon emission or absorption in the electron-photon interaction vertex. It can be ignored in our study since it plays little role in our proposed experimental techniques. 

We introduce the second quantization of the electromagnetic vector potential $\mathbf{A}$ as follows:\citep{Bruus}
\begin{equation}
\mathbf{A}(\mathbf{q}) = \sum_{\lambda=1,2} \sqrt{\frac{\hbar}{2\varepsilon_0 \omega_{\mathbf{q}} \mathcal{V} }} \mathbf{e}_{\lambda}(\mathbf{q}) (a_{\mathbf{q} \lambda} +  a^{\dag}_{-\mathbf{q} \lambda}) , \label{eqn2.1.4}
\end{equation}
where $\varepsilon_0$ is the permittivity of vacuum, $\omega_{\mathbf{q}}$ is the photon frequency, $\mathcal{V}$ is the volume for $\mathbf{A}$ to be enclosed, $\mathbf{e}_{\lambda}$ is the $\lambda$-th polarization unit vector, and $a_{\mathbf{q} \lambda}$ is the photon annihilation operator. The electron-photon interaction Eq. (\ref{eqn2.1.2}) can be expressed as 
\begin{equation}
V = \sum_{\mathbf{k}\sigma \mathbf{q}\lambda} g(\mathbf{k};\mathbf{q}\lambda) c^{\dag}_{\mathbf{k+q} \sigma} c_{\mathbf{k}\sigma} (a_{\mathbf{q}\lambda} + a^{\dag}_{\mathbf{-q} \lambda} ) , \label{eqn2.1.5}
\end{equation}
where the interaction factor $g$ is defined by
\begin{equation}
g(\mathbf{k};\mathbf{q}\lambda) = - \sqrt{\frac{\hbar}{2\varepsilon_0 \omega_{\mathbf{q}} \mathcal{V}}} \mathbf{e}_{\lambda}(\mathbf{q}) \cdot \mathbf{v}(\mathbf{k,q}) . \label{eqn2.1.6}
\end{equation}
It is noted that $g$ is a real number.

\subsection{Review of theoretical formalism for ARPES}\label{sec2.2}

The physical principle for the ARPES is the photoelectric effect. When an incident photon is absorbed by an electron in the sample material, this electron can be excited from a low-energy state into a high-energy state. If the excited electron has an enough high energy to overcome the material work function, it can escape from the sample material and emit outside to be a photoelectron. A fully defined theoretical formalism for the photon absorption and photoelectron emission in the ARPES is too complex, and in most cases, an approximate three-step model is taken. \citep{ShenRMP2003,BerglundPR1964,FeibelmanPRB1974} In this approximate model, the whole photoelectric process can be subdivided into three independent and sequential steps: the excitation of an electron in the sample bulk by the incident photon, the travel of the excited electron to the sample surface, and the emission of the photoelectron from the sample surface into vacuum. 

With an additional sudden approximation, i.e., the excited electron removes instantaneously with no post-collisional interaction with the sample material left behind,\citep{ShenRMP2003} we can introduce the following Hamiltonian to describe the photoelectric process in the ARPES:  
\begin{equation}
H = H_0 + V^{(1)}, H_0 = H_s + H_d + H_p , \label{eqn2.2.1}
\end{equation}   
where $H_s$ is the Hamiltonian of the sample electrons, $H_d$ describes the photoelectrons under the sudden approximation, and $H_p$ is the photon Hamiltonian. The electron-photon interaction $V^{(1)}$ is defined as
\begin{equation}
V^{(1)} = g\left(\mathbf{k};\mathbf{q}\lambda\right) d^{\dag}_{\mathbf{k+q} \sigma} c_{\mathbf{k}\sigma} a_{\mathbf{q}\lambda} , \label{eqn2.2.2}
\end{equation}
where $c_{\mathbf{k}\sigma}$ and $d_{\mathbf{k}\sigma}$ are the respective annihilation operators of the sample electrons and the vacuum photoelectrons. 

The emitting photoelectrons are detected by a detector, where the counting probability of this photoelectric process can be defined by 
\begin{equation}
\Gamma^{(1)} = \frac{1}{Z}\sum_{\alpha\beta} e^{-\beta E_\alpha}  \arrowvert \langle \Phi_\beta \arrowvert S^{(1)}(+\infty,-\infty) \arrowvert  \Phi_\alpha \rangle  \arrowvert^{2} , \label{eqn2.2.3}
\end{equation} 
where  
$\arrowvert  \Phi_\alpha \rangle = \arrowvert  \Psi^{s}_\alpha \rangle \otimes  \arrowvert  1_{\mathbf{q}\lambda} \rangle_p  \otimes \arrowvert  0 \rangle_d$ and  $\arrowvert  \Phi_\beta \rangle = \arrowvert  \Psi^{s}_\beta \rangle \otimes  \arrowvert 0 \rangle_p  \otimes \arrowvert  1_{\mathbf{k+q} \sigma} \rangle_d$, with the superscripts and subscripts $s$, $p$ and $d$ defined for the sample electrons, the incident photons and the photoelectrons in vacuum, respectively. The $S$-matrix $S^{(1)}(+\infty,-\infty)$ which describes the time evolution under the electron-photon interaction, is defined by
\begin{equation}
S^{(1)}(+\infty,-\infty) = -\frac{i}{\hbar} \int^{+\infty}_{-\infty} V^{(1)}_{I}(t) F(t)  dt , \label{eqn2.2.4}
\end{equation}
where $V^{(1)}_{I}(t) = e^{i H_0 t/\hbar} V^{(1)} e^{-i H_0 t/\hbar}$. The time function $F(t)$ is defined as 
\begin{equation}
F(t) = \theta(t+\Delta T /2) - \theta(t - \Delta T/2) , \label{eqn2.2.5}
\end{equation}
where $\theta(t)$ is the step function, and $\Delta T$ defines the perturbation time for the electron-photon interaction.  

It can be shown that the photoelectron counting rate $P^{(1)}\equiv \frac{\Gamma^{(1)}}{\Delta T}$ in the ARPES follows 
\begin{equation}
P^{(1)} = \frac{2\pi g^2}{\hbar}  
\frac{1}{Z}\sum_{\alpha\beta} e^{-\beta E_\alpha}  \arrowvert \langle \Psi^{s}_\beta \arrowvert c_{\mathbf{k}\sigma} \arrowvert  \Psi^{s}_\alpha \rangle \arrowvert^2 \delta(E^{(1)} + E_\beta - E_\alpha) , \label{eqn2.2.6} 
\end{equation}
where $g\equiv g\left(\mathbf{k};\mathbf{q}\lambda\right)$, $E_{\alpha}$ and $E_{\beta}$ are the eigenvalues of the eigenstates $\arrowvert  \Psi^{s}_\alpha \rangle$ and $\arrowvert  \Psi^{s}_\beta \rangle$ respectively. Here the energy $E^{(1)}$ is defined as 
\begin{equation}
E^{(1)} = \varepsilon^{(d)}_{\mathbf{k+q} \sigma} + \Phi - \hbar \omega_{\mathbf{q}} , \label{eqn2.2.7}
\end{equation}
where $\varepsilon^{(d)}$ is the energy of the photoelectrons in vacuum, and $\Phi$ is the sample material work function. It should be noted that the energy of the sample electrons is defined with respective to the Fermi energy or chemical potential. During the derivation, we have made an assumption that the time interval $\Delta T$ is large and an approximation $\frac{\sin^2(a x) }{x^2} \rightarrow \pi a \delta(x)$ when $a\rightarrow +\infty$ is used. 

We introduce the single-particle spectrum function as $A(\mathbf{k}\sigma, E) = - 2 \text{Im} G(\mathbf{k}\sigma, i\omega_n \rightarrow E + i \delta^{+})$, where $G(\mathbf{k}\sigma, i\omega_n)$ is the Fourier transformation of an imaginary-time Green's function $G(\mathbf{k}\sigma, \tau) = -\langle T_\tau c_{\mathbf{k}\sigma}(\tau) c^{\dag}_{\mathbf{k}\sigma} (0)  \rangle$ and $\delta^+$ is a positive infinitesimal, we can show that 
\begin{equation}
P^{(1)} = \frac{g^2}{\hbar} A(\mathbf{k}\sigma, E^{(1)}) n_F (E^{(1)}) , \label{eqn2.2.8} 
\end{equation}
where $n_F (E)$ is the Fermi distribution function, and the single-particle spectrum function $A(\mathbf{k}\sigma, E)$ follows 
\begin{widetext}
\begin{equation}
A(\mathbf{k}\sigma, E)=\frac{2\pi}{Z}\sum_{\alpha\beta} \left( e^{-\beta E_\alpha} + e^{-\beta E_\beta}\right)  \arrowvert \langle \Psi^{s}_\beta \arrowvert c_{\mathbf{k}\sigma} \arrowvert  \Psi^{s}_\alpha \rangle \arrowvert^2 \delta(E + E_\beta - E_\alpha) . \label{eqn2.2.9}  
\end{equation}
\end{widetext}
The photoelectron counting rate in the ARPES, Eq.(\ref{eqn2.2.8}), is the same as the Fermi's Golden-rule formula.\citep{ShenRMP2003} It shows that the detection of the angle-resolved emission of the photoelectrons can provide the momentum and energy dependent single-particle spectrum function of the sample electrons. The interaction-driven physics can then be partially investigated by the ARPES from the detected single-particle spectrum function.\citep{ShenRMP2003} 

It should be noted that in the definition of the photoelectron counting probability $\Gamma^{(1)}$ in Eq. (\ref{eqn2.2.3}), the photon states in $\vert \Phi_{\alpha} \rangle$ and $\vert \Phi_{\beta} \rangle$ are assumed to be single-photon and phonon-vacuum states, respectively. This is one approximation just for the discussion to be simple. For the realistic experimental ARPES, the photon state can be in the macroscopic coherent state or other multi-photon states. In this case, all of the above results can be similarly obtained with one additional factor to account for the redefined photon states. This approximation will be introduced in the following discussions on the photoelectric physical processes without any special notice.

\subsection{ Proposal of $\text{c}$ARPES }\label{sec2.3}

\begin{figure}[ht]
\includegraphics[width=0.9\columnwidth]{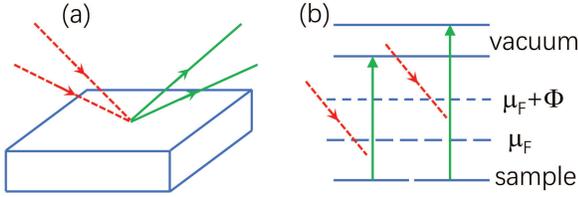} 
\caption{ (Color online) Schematic diagrams of the cARPES. In (a), the two red dashed lines represent two incident photons and the two green solid lines represent two photoelectrons. (b) The energetics of the cARPES, where the two upper blue lines with ``vacuum" denote the vacuum electron states, and the two lower blue lines with `` sample" denote the sample electron states. $\mu_F$ is the chemical potential and $\Phi$ is the work function. The line with $\mu_F+\Phi$ is the vacuum state near the sample surface with the surface effects involved. }
\label{fig2.3.1}
\end{figure}

\begin{figure}[ht]
\includegraphics[width=0.55\columnwidth]{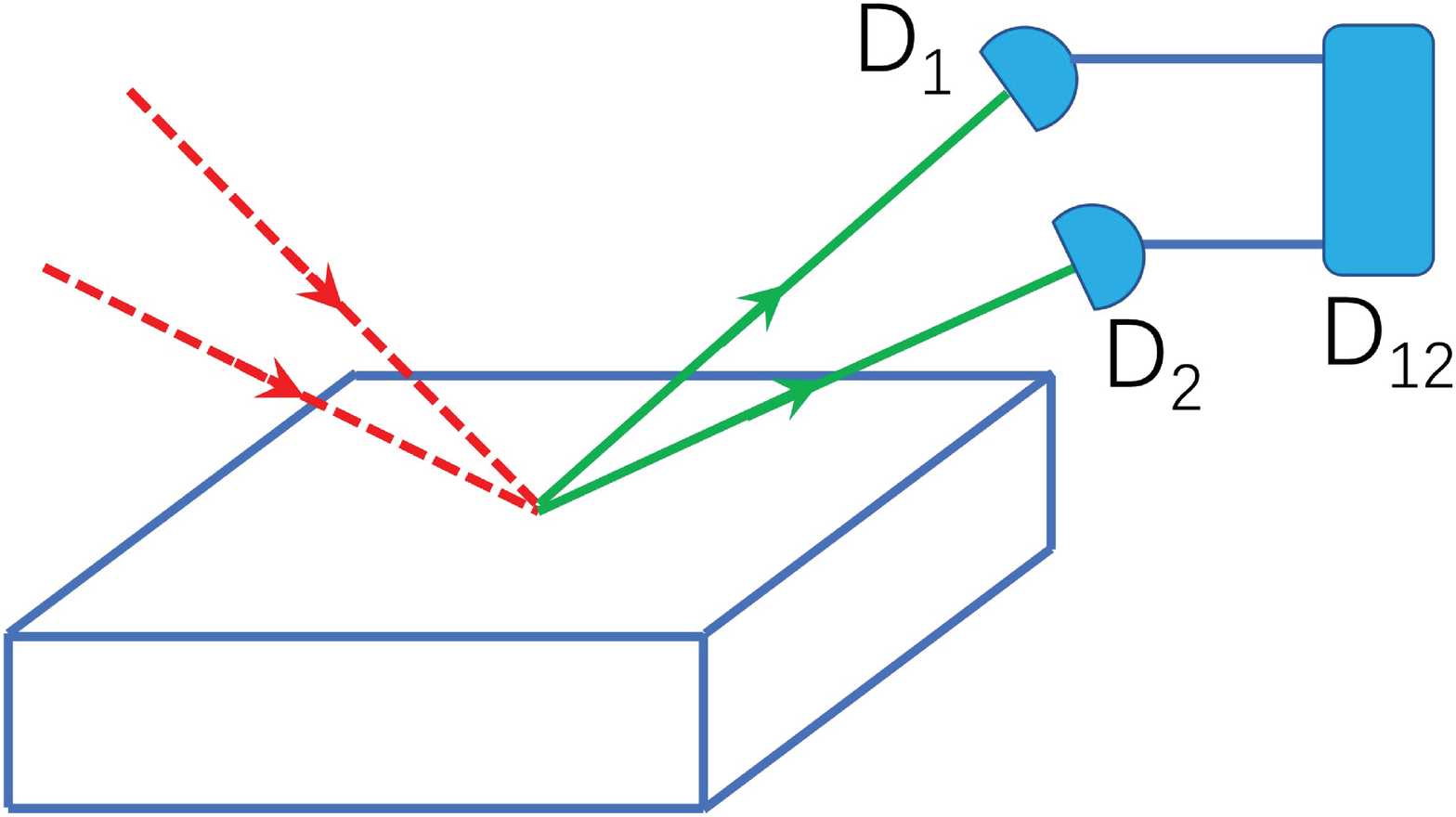} 
\caption{ (Color online) Coincidence detection of the two photoelectrons in the cARPES. D$_1$ and D$_2$ are two single-electron detectors for the photoelectrons, and D$_{12}$ is a coincidence detector which records one counting only when D$_1$ and D$_2$ each detect one photoelectron simultaneously. }
\label{fig2.3.2}
\end{figure}

A cARPES is shown schematically in Fig. \ref{fig2.3.1}. There are two photon sources which emit two photons on the sample material. These two incident photons can be absorbed by two sample electrons which are then excited into high-energy states. If their energies are high enough to overcome the material work function, the two excited electrons can escape from the sample material and emit into vacuum as two photoelectrons. A coincidence detector detects the emission of the two photoelectrons in coincidence, as schematically shown in Fig. \ref{fig2.3.2}.  

Following the discussion on the ARPES, let us establish the theoretical formalism for the coincidence detection in the cARPES. Suppose the two incident photons have momenta and polarizations $(\mathbf{q}_1,\lambda_1)$ and $(\mathbf{q}_2,\lambda_2)$. They will be absorbed by two sample electrons with $(\mathbf{k}_1,\sigma_1)$ and $(\mathbf{k}_2,\sigma_2)$, which will be excited into high-energy states and then escape into vacuum as photoelectrons with $(\mathbf{k}_1+\mathbf{q}_1,\sigma_1)$ and $(\mathbf{k}_2+\mathbf{q}_2,\sigma_2)$.     
Similar to the three-step model with the sudden approximation,\citep{ShenRMP2003,BerglundPR1964,FeibelmanPRB1974} the electron-photon interaction vertices for the two photoelectric physical processes can be defined by   
\begin{eqnarray}
V^{(2)}_1 &=&  g\left(\mathbf{k}_1;\mathbf{q}_1\lambda_1\right) d^{\dag}_{\mathbf{k}_1+\mathbf{q}_1 \sigma_1} c_{\mathbf{k}_1\sigma_1} a_{\mathbf{q}_1\lambda_1} , \notag \\
V^{(2)}_2 &=&  g\left(\mathbf{k}_2;\mathbf{q}_2\lambda_2\right) d^{\dag}_{\mathbf{k}_2+\mathbf{q}_2 \sigma_2} c_{\mathbf{k}_2\sigma_2} a_{\mathbf{q}_2\lambda_2} . \notag 
\end{eqnarray}

The coincidence probability recorded by the coincidence detector in the cARPES is defined by 
\begin{equation}
\Gamma^{(2)} = \frac{1}{Z}\sum_{\alpha\beta} e^{-\beta E_\alpha}  \arrowvert \langle \Phi_\beta \arrowvert S^{(2)}(+\infty,-\infty) \arrowvert  \Phi_\alpha \rangle  \arrowvert^{2} , \label{2.3.1}
\end{equation} 
where  
$\arrowvert  \Phi_\alpha \rangle = \arrowvert  \Psi^{s}_\alpha \rangle \otimes  \arrowvert  1_{\mathbf{q}_1\lambda_1} 1_{\mathbf{q}_2\lambda_2} \rangle_p  \otimes \arrowvert  0 \rangle_d$ and  $\arrowvert  \Phi_\beta \rangle = \arrowvert  \Psi^{s}_\beta \rangle \otimes  \arrowvert  0\rangle_p  \otimes \arrowvert  1_{\mathbf{k}_1+\mathbf{q}_1 \sigma_1} 1_{\mathbf{k}_2+\mathbf{q}_2 \sigma_2} \rangle_d$. The relevant $S$ matrix is defined as
\begin{widetext}
\begin{equation}
S^{(2)}(+\infty,-\infty) = \left(-\frac{i}{\hbar}\right)^2 \iint^{+\infty}_{-\infty} T_{t} \lbrack V^{(2)}_{2,I}(t_2) V^{(2)}_{1,I}(t_1) \rbrack F(t_2) F(t_1)  dt_2 dt_1 , \label{eqn2.3.2}
\end{equation}
where $V^{(2)}_{i,I}(t) = e^{i H_0 t/\hbar} V^{(2)}_i e^{-i H_0 t/\hbar}$ with $H_0$ defined in Eq. (\ref{eqn2.2.1}) and $T_{t}$ is the time ordering operator. The time function $F(t)$ is given in Eq. (\ref{eqn2.2.5}). It is shown that the coincidence probability of the cARPES follows
\begin{equation}
\Gamma^{(2)} =  \frac{ \left( g_1 g_2 \right)^2 }{\hbar^4}\frac{1}{Z}\sum_{\alpha\beta} e^{-\beta E_\alpha} \bigg\vert \iint^{+\infty}_{-\infty} \phi^{(2)}_{\alpha\beta}\left( \mathbf{k}_1\sigma_1 t_1; \mathbf{k}_2\sigma_2 t_2 \right) e^{i( E_1^{(2)} t_1 + E_2^{(2)} t_2)/\hbar} F(t_2) F(t_1)  dt_2 dt_1 \bigg\vert^2 , \label{eqn2.3.3}
\end{equation} 
where $\phi^{(2)}_{\alpha\beta}\left( \mathbf{k}_1\sigma_1 t_1; \mathbf{k}_2\sigma_2 t_2 \right)$ is a Bethe-Salpeter wave function\citep{GellmanLowBS1951,SalpeterBethe1951} defined in the particle-particle channel as 
\begin{equation}
\phi^{(2)}_{\alpha\beta}\left( \mathbf{k}_1\sigma_1 t_1; \mathbf{k}_2\sigma_2 t_2 \right) = \langle \Psi^{s}_{\beta} \arrowvert T_t c_{\mathbf{k}_2 \sigma_2} \left(t_2\right) c_{\mathbf{k}_1 \sigma_1} \left(t_1\right) \arrowvert \Psi^{s}_{\alpha} \rangle . \label{eqn2.3.4}
\end{equation}
In Eq. (\ref{eqn2.3.3}), $g_1 \equiv g(\mathbf{k}_1; \mathbf{q}_1\lambda_1)$ and $g_2 \equiv g(\mathbf{k}_2; \mathbf{q}_2\lambda_2)$, and the transfer energies $ E^{(2)}_1$ and $ E^{(2)}_2$ are defined as
\begin{eqnarray}
&& E^{(2)}_1 = \varepsilon^{(d)}_{\mathbf{k}_1+\mathbf{q}_1 \sigma_1} + \Phi - \hbar \omega_{\mathbf{q}_1} ,  E^{(2)}_2 = \varepsilon^{(d)}_{\mathbf{k}_2+\mathbf{q}_2 \sigma_2} + \Phi - \hbar \omega_{\mathbf{q}_2} . \label{eqn2.3.5} 
\end{eqnarray}

The time integrals in the coincidence probability $\Gamma^{(2)}$ show that it involves a Fourier-transformation-like structure of the Bethe-Salpeter wave function. This can be explicitly shown in the limit $\Delta T \rightarrow +\infty$, where  the coincidence probability of the cARPES in Eq. (\ref{eqn2.3.3}) becomes 
\begin{equation}
\Gamma^{(2)} = \frac{\left( g_1 g_2 \right)^2 }{ \hbar^4} \frac{1}{Z} \sum_{\alpha\beta} e^{-\beta E_{\alpha}} \big\vert \phi^{(2)}_{\alpha\beta}\left(\mathbf{k}_1\sigma_1, \mathbf{k}_2\sigma_2; \Omega_c, \omega_r \right) \big\vert^2 . \label{eqn2.3.6}
\end{equation}
Here we have introduced the Fourier transformations,
\begin{eqnarray}
&&\phi^{(2)}_{\alpha\beta}\left(\mathbf{k}_1\sigma_1, \mathbf{k}_2\sigma_2; t_c, t_r \right) = \frac{1}{\left(2 \pi \right)^2} \iint_{-\infty}^{+\infty} \phi^{(2)}_{\alpha\beta}\left( \mathbf{k}_1\sigma_1, \mathbf{k}_2\sigma_2; \Omega, \omega \right) e^{-i\Omega t_c - i\omega t_r} d\Omega d\omega , \notag \\
&&\phi^{(2)}_{\alpha\beta}\left(\mathbf{k}_1\sigma_1, \mathbf{k}_2\sigma_2 ; \Omega, \omega \right) =  \iint_{-\infty}^{+\infty} \phi^{(2)}_{\alpha\beta}\left(\mathbf{k}_1\sigma_1, \mathbf{k}_2\sigma_2 ; t_c, t_r \right) e^{i\Omega t_c + i\omega t_r} d t_c d t_r , \notag
\end{eqnarray}
where $\phi^{(2)}_{\alpha\beta}\left( \mathbf{k}_1\sigma_1, \mathbf{k}_2\sigma_2; t_c, t_r \right) = \phi^{(2)}_{\alpha\beta}\left( \mathbf{k}_1\sigma_1 t_1; \mathbf{k}_2\sigma_2 t_2 \right)$ with the center-of-mass time $t_c$ and the relative time $t_r$ defined by 
\begin{equation}
t_c = \frac{1}{2} \left( t_1 + t_2 \right), t_r = t_2 - t_1 . \label{eqn2.3.7} 
\end{equation}
The center-of-mass frequency $\Omega_c$ and the inner-pair relative frequency $\omega_r$ in Eq. (\ref{eqn2.3.6}) are set as 
\begin{equation}
\Omega_c=E^{(2)}/\hbar, \omega_r =\mathcal{E}^{(2)}/\hbar , \label{eqn2.3.8} 
\end{equation}
with the transfer energies $E^{(2)}$ and $\mathcal{E}^{(2)}$ defined as
\begin{equation}
E^{(2)} = E_1^{(2)} + E_2^{(2)} , \mathcal{E}^{(2)} = \frac{1}{2} ( E_2^{(2)} - E_1^{(2)} ) . \label{eqn2.3.9}
\end{equation}

The coincidence probability of the cARPES in the approximate limit $\Delta T \rightarrow +\infty$, Eq. (\ref{eqn2.3.6}), shows that it provides directly the information on the frequency Bethe-Salpeter wave function. The frequency Bethe-Salpeter wave function has a general form:  
\begin{equation}
\phi^{(2)}_{\alpha\beta}\left(\mathbf{k}_1\sigma_1, \mathbf{k}_2\sigma_2; \Omega, \omega \right)  = 2\pi \delta \left[\Omega + \left( E_{\beta} - E_{\alpha}\right)/\hbar \right] \phi^{(2)}_{\alpha\beta}\left(\mathbf{k}_1\sigma_1, \mathbf{k}_2\sigma_2; \omega \right) , \label{eqn2.3.10} 
\end{equation} 
where $\phi^{(2)}_{\alpha\beta}\left(\mathbf{k}_1\sigma_1, \mathbf{k}_2\sigma_2; \omega \right)$ follows
\begin{equation}
\phi^{(2)}_{\alpha\beta}\left(\mathbf{k}_1\sigma_1, \mathbf{k}_2\sigma_2; \omega \right) =\sum_{\gamma} \left[ \frac{ i \langle\Psi^{s}_{\beta} \vert c_{\mathbf{k}_2 \sigma_2} \vert \Psi^s_{\gamma} \rangle \langle \Psi^s_{\gamma} \vert  c_{\mathbf{k}_1 \sigma_1} \vert \Psi^s_{\alpha} \rangle} {\omega + i\delta^+ + (E_{\alpha} + E_{\beta} - 2 E_{\gamma} )/2\hbar} + \frac{ i \langle\Psi^{s}_{\beta} \vert c_{\mathbf{k}_1 \sigma_1} \vert \Psi^s_{\gamma} \rangle \langle \Psi^s_{\gamma} \vert  c_{\mathbf{k}_2 \sigma_2} \vert \Psi^s_{\alpha} \rangle} {\omega - i\delta^+ - (E_{\alpha} + E_{\beta} - 2 E_{\gamma} )/2\hbar} \right] . \label{eqn2.3.11}
\end{equation}
The two-particle Bethe-Salpeter wave function for the cARPES describes the physics of the sample electrons when two electrons are annihilated in time ordering, thus it describes the particle-particle pair dynamical physics of the sample electrons (more exactly, hole-hole pair). The frequency Bethe-Salpeter wave function involves the following physics: (1) The pair center-of-mass dynamical physics of the sample electrons described by the $\delta$-function, $\delta \left[\Omega + \left( E_{\beta} - E_{\alpha}\right)/\hbar \right]$, which shows the energy transfer conservation for the pair center-of-mass degrees of freedom; (2) The inner-pair dynamical physics of the sample electrons described by 
$\phi^{(2)}_{\alpha\beta}\left(\mathbf{k}_1\sigma_1, \mathbf{k}_2\sigma_2; \omega \right)$, which shows the propagatorlike resonance structures, peaked at $\omega = \pm (E_{\alpha} + E_{\beta} - 2 E_{\gamma} )/2\hbar$ with the weights defined by $\langle\Psi^{s}_{\beta} \vert c_{\mathbf{k}_2 \sigma_2} \vert \Psi^s_{\gamma} \rangle \langle \Psi^s_{\gamma} \vert  c_{\mathbf{k}_1 \sigma_1} \vert \Psi^s_{\alpha} \rangle$ and $\langle\Psi^{s}_{\beta} \vert c_{\mathbf{k}_1 \sigma_1} \vert \Psi^s_{\gamma} \rangle \langle \Psi^s_{\gamma} \vert  c_{\mathbf{k}_2 \sigma_2} \vert \Psi^s_{\alpha} \rangle$. When the two electrons annihilated are independent without correlations such as in the free Fermi gas, $\phi^{(2)}_{\alpha\beta}\left(\mathbf{k}_1\sigma_1, \mathbf{k}_2\sigma_2; \omega \right)$ has a behavior of the two-particle spectrum function as $2\pi \langle\Psi^{s}_{\beta} \vert c_{\mathbf{k}_2 \sigma_2} c_{\mathbf{k}_1 \sigma_1} \vert \Psi^s_{\alpha} \rangle \delta \left[ \omega - \left(\varepsilon_{\mathbf{k}_2} - \varepsilon_{\mathbf{k}_1}\right)/2\hbar \right]$, where $\varepsilon_{\mathbf{k}_{1}}$ and $\varepsilon_{\mathbf{k}_{2}}$ are the energies of the two free electrons.

For one particle-particle pair with $\left(\mathbf{k}_1\sigma_1, \mathbf{k}_2\sigma_2\right)$, when scan the energies $E^{(2)}_1$ and $E^{(2)}_2$ and thus $E^{(2)}$ and $\mathcal{E}^{(2)}$, the momentum and energy dependent Bethe-Salpeter wave function in its absolute value can be obtained by the cARPES, as shown by Eq. (\ref{eqn2.3.6}). Therefore, the cARPES can provide the momentum and energy dependent particle-particle pair dynamical physics of the sample electrons.  Moreover, the center-of-mass and the inner-pair relative dynamics of the sample electrons can also be resolved by the cARPES. One more interesting is that if the spin configurations of the photoelectrons can be detected, the spin magnetic properties can also be studied by the cARPES. In this case, the cARPES can provide the momentum-energy-spin resolved Bethe-Salpeter wave function and the relevant two-particle correlations in the particle-particle channel. These discussions show that the cARPES is one potential technique to study the Cooper-pair physics in the unconventional superconductors, especially the time-retarded dynamical physics which is deeply relevant to the microscopic pairing mechanism.   

For the finite but large $\Delta T$, the coincidence probability of the cARPES can be expressed as
\begin{equation}
\Gamma^{(2)} = \frac{\left( g_1 g_2 \right)^2 }{(2\pi \hbar)^4} \frac{1}{Z}\sum_{\alpha\beta} e^{-\beta E_{\alpha}} \bigg\vert \iint^{+\infty}_{-\infty} d\Omega	 d\omega \phi^{(2)}_{\alpha\beta}\left(\mathbf{k}_1\sigma_1, \mathbf{k}_2\sigma_2; \Omega, \omega \right) Y(\Omega,\omega) \bigg\vert^2 , \label{eqn2.3.12}
\end{equation}
where $Y(\Omega,\omega)$ is defined by
\begin{equation}
Y(\Omega,\omega)= \frac{2 \sin  \lbrack ( \Omega - E^{(2)} /\hbar )\Delta T/2 \rbrack }{\Omega-E^{(2)}/\hbar } \cdot \frac{2 \sin  \lbrack ( \omega - \mathcal{E}^{(2)} /\hbar )\Delta T/2 \rbrack }{\omega - \mathcal{E}^{(2)}/\hbar } . \label{eqn2.3.13}
\end{equation}
The coincidence probability of the cARPES, Eq. (\ref{eqn2.3.12}), can be regarded to be relevant to one finite-$\Delta T$ restricted Fourier-transformation of the Bethe-Salpeter wave function. 
Here we have used the large $\Delta T$ approximation that $\int_{-\Delta T/2}^{\Delta T/2}d t_2 \int_{-\Delta T/2}^{\Delta T/2}d t_1 \rightarrow \int_{-\Delta T/2}^{\Delta T/2}d t_c \int_{-\Delta T/2}^{\Delta T/2}d t_r $. Since the frequency Bethe-Salpeter wave function has the general form as Eq. (\ref{eqn2.3.10}), the coincidence probability of the cARPES can be calculated from the following expression: 
\begin{equation}
\Gamma^{(2)} = \frac{\left( g_1 g_2 \right)^2 \Delta T}{2\pi \hbar^3} \frac{1}{Z}\sum_{\alpha\beta} e^{-\beta E_{\alpha}} \delta  ( E^{(2)} + E_{\beta} - E_{\alpha} ) I^{(2)}_{\alpha\beta}\left(\mathbf{k}_1\sigma_1, \mathbf{k}_2\sigma_2  \right) , \label{eqn2.3.14}
\end{equation}
where $I^{(2)}_{\alpha\beta}\left(\mathbf{k}_1\sigma_1, \mathbf{k}_2\sigma_2  \right)$ is defined by 
\begin{equation}
I^{(2)}_{\alpha\beta}\left(\mathbf{k}_1\sigma_1, \mathbf{k}_2\sigma_2  \right) = \bigg\vert \int^{+\infty}_{-\infty} d \omega \phi^{(2)}_{\alpha\beta}\left(\mathbf{k}_1\sigma_1, \mathbf{k}_2\sigma_2; \omega \right) \frac{2 \sin  \lbrack ( \omega - \mathcal{E}^{(2)} /\hbar )\Delta T/2 \rbrack }{\omega - \mathcal{E}^{(2)}/\hbar } \bigg\vert^2 . \label{eqn2.3.15}
\end{equation}
\end{widetext}

Let us now give a remark on the experimental installation of the cARPES. In our above proposal for the cARPES, the two incident photons are assumed to come from two photon sources. Since each beam from one source will lead to the photoelectron emission in all of the different angles, to distinguish correctly which is the corresponding emitting photoelectron from one given incident beam needs more experimental tricks. In a realistic experimental installation, one single photon source can emit two photons which can lead to the following photoelectric effects for the cARPES. In this single-source cARPES, the two electron-photon interaction vertices can be similarly defined with the two incident photons having the same momentum and polarization $(\mathbf{q},\lambda)$. All of the above results on the coincidence probability of the two-source cARPES can be similarly derived for the single-source cARPES, with only the substitution of $(\mathbf{q}_1,\lambda_1)=(\mathbf{q}_2,\lambda_2)=(\mathbf{q},\lambda)$. Thus, a simple experimental installation of the cARPES can be built upon an installation of the ARPES with one additional coincidence detector.    

Recently, one photoemission technique, double photoemission spectroscopy, has been developed to study the electron correlations.\citep{BerakdarPRB1998,ChiangPRL2017,AliaevSS2018} In this double photoemission spectroscopy, one photon is absorbed which excites the sample electrons into high-energy states. The disturbed sample electrons emit two electrons which are detected in coincidence. As microscopically one single photon can only excite one single electron, the one-photon-absorption double photoemission spectroscopy will involve subsequential intermediate excited states which contribute to the two-electron emission. It is different in principle to the proposed cARPES in the study of the two-particle correlations.

\section{ $\text{c}$ARPES for free Fermi gas and superconducting state }\label{sec3}

We will study the cARPES spectra of a free Fermi gas and a BCS superconducting state in this section. 

\subsection{ Free Fermi gas }\label{sec3.1}

A free Fermi gas has a Hamiltonian 
\begin{equation}
H = \sum_{\mathbf{k}\sigma} \varepsilon_{\mathbf{k}} c^{\dag}_{\mathbf{k}\sigma} c_{\mathbf{k}\sigma} , \label{eqn3.1.1}
\end{equation}
where the chemical potential has been included in $\varepsilon_{\mathbf{k}}$. It can be easily shown that the two-particle Bethe-Salpeter wave function of the free Fermi gas follows 
\begin{widetext}
\begin{equation}
\phi^{(2)}_{\alpha\beta}(\mathbf{k}_1\sigma_1,\mathbf{k}_2\sigma_2; \Omega, \omega) = 2\pi \delta \left[ \Omega - \left(\varepsilon_{\mathbf{k}_1} + \varepsilon_{\mathbf{k}_2}\right)/\hbar \right] \phi^{(2)}_{\alpha\beta}(\mathbf{k}_1\sigma_1,\mathbf{k}_2\sigma_2; \omega) , \label{eqn3.1.2}
\end{equation}
where $\phi^{(2)}_{\alpha\beta}(\mathbf{k}_1\sigma_1,\mathbf{k}_2\sigma_2; \omega)$ follows
\begin{equation}
\phi^{(2)}_{\alpha\beta}(\mathbf{k}_1\sigma_1,\mathbf{k}_2\sigma_2; \omega) =  2\pi \delta \left[ \omega - \left(\varepsilon_{\mathbf{k}_2} - \varepsilon_{\mathbf{k}_1}\right)/2\hbar \right] \delta(n_{\mathbf{k}_1 \sigma_1}-1) \delta(n_{\mathbf{k}_2 \sigma_2}-1)  \label{eqn3.1.3}
\end{equation}
when $\arrowvert\Psi^{s}_\beta\rangle = c_{\mathbf{k}_2 \sigma_2} c_{\mathbf{k}_1 \sigma_1} \arrowvert \Psi^{s}_\alpha\rangle$, and it is zero for the other cases. Here $n_{\mathbf{k}_1 \sigma_1}, n_{\mathbf{k}_2 \sigma_2}=0,1$, which define the occupation of the free Fermi particles in the state $\arrowvert \Psi^{s}_\alpha\rangle$.  The coincidence probability of the cARPES for the free Fermi gas is calculated from Eq. (\ref{eqn2.3.12}) or (\ref{eqn2.3.14}) as
\begin{equation}
\Gamma^{(2)} = \frac{4\pi^2 \left( g_1 g_2 \right)^2 \Delta T^2}{\hbar^2} \delta (E^{(2)} - \varepsilon_{\mathbf{k}_1} - \varepsilon_{\mathbf{k}_2} ) \delta [ \mathcal{E}^{(2)} -(\varepsilon_{\mathbf{k}_2} - \varepsilon_{\mathbf{k}_1})/2 ] n_F ( \varepsilon_{\mathbf{k}_1}) n_F ( \varepsilon_{\mathbf{k}_2})  . \label{eqn3.1.4} 
\end{equation}
Obviously, the coincidence probability of the cARPES shows the information of the dynamical frequency Bethe-Salpeter wave function $\phi^{(2)}_{\alpha\beta}(\mathbf{k}_1\sigma_1,\mathbf{k}_2\sigma_2; \Omega, \omega)$.
At zero temperature, the coincidence probability of the cARPES behaves as 
\begin{equation}
\Gamma^{(2)} = \frac{4\pi^2 \left( g_1 g_2 \right)^2 \Delta T^2}{\hbar^2} \delta (E^{(2)} - \varepsilon_{\mathbf{k}_1} - \varepsilon_{\mathbf{k}_2} ) \delta [ \mathcal{E}^{(2)} -(\varepsilon_{\mathbf{k}_2} - \varepsilon_{\mathbf{k}_1})/2 ] \theta( - \varepsilon_{\mathbf{k}_1}) \theta(  - \varepsilon_{\mathbf{k}_2}) . \label{eqn3.1.5} 
\end{equation}
\end{widetext}

Since the single-particle ARPES spectrum of the free Fermi gas follows $\Gamma^{(1)} = \frac{2\pi g^2 \Delta T}{\hbar} \delta ( E^{(1)} - \varepsilon_{\mathbf{k}} ) n_F (\varepsilon_{\mathbf{k}})$, we have the following relation:
\begin{equation}
\Gamma^{(2)} = \Gamma^{(1)}\left( g_1 \right) \Gamma^{(1)}\left( g_2 \right) , \label{eqn3.1.6}  
\end{equation}
where $\Gamma^{(1)}\left( g_1 \right)$ and $\Gamma^{(1)}\left( g_2 \right)$ are the two respective single-particle ARPES counting probabilities of the two photoelectric processes in the cARPES. This relation shows that the coincidence probability of the cARPES for the Fermi free gas is trivial with product contribution from two independent photoelectric processes. This is consistent with the fact that the free Fermi gas has only single-particle physics without two-particle correlations.

\subsection{ Superconducting state }\label{sec3.2}

Let us consider a superconducting state with spin singlet pairing. In a mean-field approximation, the superconducting state can be described by a BCS mean-field Hamiltonian
\begin{equation}
H_{BCS} = \sum_{\mathbf{k}\sigma} \varepsilon_{\mathbf{k}} c^{\dag}_{\mathbf{k}\sigma} c_{\mathbf{k}\sigma} + \sum_{\mathbf{k}}\left( \Delta^{\ast}_{\mathbf{k}} c_{-\mathbf{k}\downarrow} c_{\mathbf{k}\uparrow} + \Delta_{\mathbf{k}} c^{\dag}_{\mathbf{k}\uparrow} c^{\dag}_{-\mathbf{k}\downarrow} \right) , \label{eqn3.2.1}
\end{equation}
where $\Delta_{\mathbf{k}}=\vert \Delta_{\mathbf{k}} \vert e^{i\theta_{\mathbf{k}}} $ is a $\mathbf{k}$-dependent gap function. We introduce the Bogoliubov transformations 
\begin{equation}
\left(\begin{array}{c}
\alpha_{\mathbf{k}\uparrow} \\
\alpha^{\dag}_{-\mathbf{k}\downarrow} 
\end{array} \right) 
= \left( \begin{array}{cc}
u_{\mathbf{k}} & v_{\mathbf{k}} \\
- v^{\ast}_{\mathbf{k}} & u_{\mathbf{k}} 
\end{array} \right) 
\left(\begin{array}{c}
c_{\mathbf{k}\uparrow} \\
c^{\dag}_{-\mathbf{k}\downarrow}
\end{array} \right) , \label{eqn3.2.2}
\end{equation} 
where $u_{\mathbf{k}}$ and $v_{\mathbf{k}}$ are defined by
\begin{equation}
u_{\mathbf{k}} = \sqrt{\frac{1}{2}\left(1+\frac{\varepsilon_{\mathbf{k}}}{E_{\mathbf{k}}}\right)} ,
v_{\mathbf{k}} = e^{i\theta_{\mathbf{k}}} \sqrt{\frac{1}{2}\left(1-\frac{\varepsilon_{\mathbf{k}}}{E_{\mathbf{k}}}\right)}  , \label{eqn3.2.3}
\end{equation} 
the BCS Hamiltonian can be diagonalized into the form
\begin{equation}
H_{BCS} = \sum_{\mathbf{k}} E_{\mathbf{k}} \left( \alpha^{\dag}_{\mathbf{k}\uparrow}\alpha_{\mathbf{k}\uparrow} + \alpha^{\dag}_{-\mathbf{k}\downarrow} \alpha_{-\mathbf{k}\downarrow} \right)  \label{eqn3.2.4}
\end{equation}
with $E_{\mathbf{k}} = \sqrt{\varepsilon^{2}_{\mathbf{k}}+ \vert \Delta_{\mathbf{k}} \vert^2 }$.

Let us study the particle-particle Bethe-Salpeter wave function $\phi^{(2)}_{\alpha\beta}$ for a Cooper pair with $\left( \mathbf{k}\uparrow,-\mathbf{k}\downarrow \right)$. Defining $\mathbf{k}_1=\mathbf{k},\sigma_1=\uparrow, \mathbf{k}_2=-\mathbf{k},\sigma_2=\downarrow$, $\phi^{(2)}_{\alpha\beta}$ is shown to follow
\begin{equation}
\phi^{(2)}_{\alpha\beta}\left( \mathbf{k}\uparrow,-\mathbf{k}\downarrow; \Omega,\omega \right) = \sum_{i=1}^{3} \phi^{(2)}_{\alpha\beta,i}\left( \mathbf{k}\uparrow,-\mathbf{k}\downarrow; \Omega,\omega \right) , \label{eqn3.2.5}
\end{equation}
where 
\begin{eqnarray}
&&\phi^{(2)}_{\alpha\beta,1}\left( \mathbf{k}\uparrow,-\mathbf{k}\downarrow; \Omega,\omega \right) = 2\pi \delta \left(\Omega\right) \phi^{(2)}_{\alpha\beta,1}\left( \mathbf{k}\uparrow,-\mathbf{k}\downarrow; \omega \right), \notag \\
&&\phi^{(2)}_{\alpha\beta,2}\left( \mathbf{k}\uparrow,-\mathbf{k}\downarrow; \Omega,\omega \right) = 2\pi \delta \left(\Omega+2E_{\mathbf{k}}\right) \phi^{(2)}_{\alpha\beta,2}\left( \mathbf{k}\uparrow,-\mathbf{k}\downarrow; \omega \right),\notag \\
&&\phi^{(2)}_{\alpha\beta,3}\left( \mathbf{k}\uparrow,-\mathbf{k}\downarrow; \Omega,\omega \right) = 2\pi \delta \left(\Omega - 2E_{\mathbf{k}}\right) \phi^{(2)}_{\alpha\beta,3}\left( \mathbf{k}\uparrow,-\mathbf{k}\downarrow; \omega \right) .\notag 
\end{eqnarray}
The three Bethe-Salpether wave functions with only relative time dynamics follow
\begin{widetext}
\begin{eqnarray}
&&\phi^{(2)}_{\alpha\beta,1}\left( \mathbf{k}\uparrow,-\mathbf{k}\downarrow; \omega \right) = i \left( u_{\mathbf{k}} v_{\mathbf{k}} \right) \left( \frac{n^{\alpha}_{\mathbf{k}\uparrow}}{\omega + E_{\mathbf{k}}/\hbar + i\delta^{+}}  + \frac{1-n^{\alpha}_{\mathbf{k}\uparrow}}{\omega + E_{\mathbf{k}}/\hbar - i\delta^{+}} - \frac{1-n^{\alpha}_{-\mathbf{k}\downarrow}}{\omega - E_{\mathbf{k}}/\hbar + i\delta^{+}} -\frac{n^{\alpha}_{-\mathbf{k}\downarrow}}{\omega - E_{\mathbf{k}}/\hbar - i\delta^{+}}  \right) , \notag \\
&& \phi^{(2)}_{\alpha\beta,2}\left( \mathbf{k}\uparrow,-\mathbf{k}\downarrow; \omega \right) = -2\pi \delta(\omega) v^2_{\mathbf{k}} \delta(n^{\alpha}_{\mathbf{k}\uparrow}) \delta(n^{\alpha}_{-\mathbf{k}\downarrow}) , \label{eqn3.2.7} \\
&& \phi^{(2)}_{\alpha\beta,3}\left( \mathbf{k}\uparrow,-\mathbf{k}\downarrow; \omega \right) = 2\pi \delta(\omega) u^2_{\mathbf{k}} \delta(n^{\alpha}_{\mathbf{k}\uparrow}-1) \delta(n^{\alpha}_{-\mathbf{k}\downarrow}-1) , \notag
\end{eqnarray}
where $n^{\alpha}_{\mathbf{k}\sigma} = 0, 1$ which describe the occupation of the Bogoliubov quasiparticles in the state $\arrowvert \Psi^{s}_\alpha \rangle$, and $\arrowvert \Psi^{s}_\beta \rangle = \arrowvert \Psi^{s}_\alpha \rangle$ in $\phi^{(2)}_{\alpha\beta,1}$, $\arrowvert \Psi^{s}_\beta \rangle = \alpha^{\dag}_{\mathbf{k}\uparrow} \alpha^{\dag}_{-\mathbf{k}\downarrow}  \arrowvert \Psi^{s}_\alpha \rangle$ in $\phi^{(2)}_{\alpha\beta,2}$, $\arrowvert \Psi^{s}_\beta \rangle = \alpha_{-\mathbf{k}\downarrow} \alpha_{\mathbf{k}\uparrow}  \arrowvert \Psi^{s}_\alpha \rangle$ in $\phi^{(2)}_{\alpha\beta,3}$.

The coincidence probability of the cARPES for the BCS superconducting state can be calculated from Eq. (\ref{eqn2.3.12}) or (\ref{eqn2.3.14}), which follows
\begin{equation}
\Gamma^{(2)} = \Gamma^{(2)}_1 + \Gamma^{(2)}_2 + \Gamma^{(2)}_3 , \label{eqn3.2.8}
\end{equation}
where the three contributions are defined as
\begin{eqnarray}
&& \Gamma^{(2)}_1 = \frac{2\pi^2 \left( g_1 g_2 \right)^2 \Delta T^2}{\hbar^2} \arrowvert u_{\mathbf{k}} v_{\mathbf{k}} \arrowvert^2 \delta ( E^{(2)} ) \big\lbrack \delta ( \mathcal{E}^{(2)} +E_{\mathbf{k}} ) + \delta ( \mathcal{E}^{(2)} - E_{\mathbf{k}} ) \big\rbrack  , \notag \\
&& \Gamma^{(2)}_2 = \frac{4\pi^2 \left( g_1 g_2 \right)^2 \Delta T^2 }{\hbar^2} \arrowvert v_{\mathbf{k}} \arrowvert^4 \delta ( E^{(2)} + 2 E_{\mathbf{k}} ) \delta ( \mathcal{E}^{(2)} ) n^{2}_F ( - E_{\mathbf{k}} ) , \label{eqn3.2.9} \\
&& \Gamma^{(2)}_3 = \frac{4\pi^2 \left( g_1 g_2 \right)^2 \Delta T^2 }{\hbar^2} \arrowvert u_{\mathbf{k}} \arrowvert^4 \delta ( E^{(2)} - 2 E_{\mathbf{k}} ) \delta ( \mathcal{E}^{(2)} ) n^{2}_F ( E_{\mathbf{k}} ) . \notag
\end{eqnarray}
\end{widetext}
The first term $\Gamma^{(2)}_1$ comes from the contribution of $\phi^{(2)}_{\alpha\beta,1}$, which is the well-known anomalous Green's function and describes the propagators of the single Bogoliubov quasiparticles $\langle \Psi^s_{\alpha} \vert T_t \alpha^{\dag}_{\mathbf{k}\uparrow} (t_2) \alpha_{\mathbf{k}\uparrow} (t_1) \vert \Psi^s_{\alpha} \rangle$ and $\langle \Psi^s_{\alpha} \vert T_t \alpha_{-\mathbf{k}\downarrow} (t_2) \alpha^{\dag}_{-\mathbf{k}\downarrow} (t_1) \vert \Psi^s_{\alpha} \rangle$ with additional factors $\pm u_{\mathbf{k}} v_{\mathbf{k}}$. It has two resonance peak structures at $\omega_r = \pm E_{\mathbf{k}}$ in the inner-pair channel with the center-of-mass transfer energy $\Omega_c=0$. The second term $\Gamma^{(2)}_2$ comes from the contribution of $\phi^{(2)}_{\alpha\beta,2}$. It has a wave function weight factor $\arrowvert v_{\mathbf{k}} \arrowvert^4 n^{2}_F ( - E_{\mathbf{k}} )$ which comes from $v_{\mathbf{k}}^2 \langle \Psi^{s}_\beta \arrowvert \alpha^{\dag}_{\mathbf{k}\uparrow} \alpha^{\dag}_{-\mathbf{k}\downarrow} \arrowvert \Psi^{s}_\alpha \rangle$, and shows the transfer energy of the center-of-mass of the Cooper pair finite $\Omega_c = -2E_{\mathbf{k}}$ and the inner-pair relative dynamics with a resonance peak at $\omega_r =0$. The third term $\Gamma^{(2)}_3$ has a similar behavior to $\Gamma^{(2)}_2$. It involves a wave function distribution factor $\arrowvert u_{\mathbf{k}} \arrowvert^4 n^{2}_F ( E_{\mathbf{k}} )$ which comes from $u_{\mathbf{k}}^2 \langle \Psi^{s}_\beta \arrowvert \alpha_{-\mathbf{k}\downarrow} \alpha_{\mathbf{k}\uparrow} \arrowvert \Psi^{s}_\alpha \rangle$, and shows a resonance peak at $\omega_r=0$ in the inner-pair channel with the the center-of-mass transfer energy finite $\Omega_c = 2 E_{\mathbf{k}}$. The first term $\Gamma^{(2)}_1$ is intrinsic to the macroscopic coherent superconducting state and proportional to the square of the gap function as $u_{\mathbf{k}} v_{\mathbf{k}}=\frac{\Delta_{\mathbf{k}}}{2 E_{\mathbf{k}}}$. It reduces to zero in the normal state with zero superconducting gap, where the coincidence probability $\Gamma^{(2)}$ shows the behavior of two free electrons the same as the formula [Eq. (\ref{eqn3.1.4})] of the free Fermi gas. 

The coherent superconducting ground state $\vert \Psi_{BCS} \rangle = C e^{\sum_{\mathbf{k}}\psi_{\mathbf{k}} c_{-\mathbf{k}\downarrow}^{\dag} c_{\mathbf{k}\uparrow}^{\dag}}\vert 0 \rangle=\prod_{\mathbf{k}} ( u_{\mathbf{k}} + v_{\mathbf{k}}  c_{-\mathbf{k}\downarrow}^{\dag} c_{\mathbf{k}\uparrow}^{\dag} ) \vert 0 \rangle$, where $\psi_{\mathbf{k}}$ is the inner-Cooper-pair wave function, $u_{\mathbf{k}}=\frac{1}{\sqrt{1+\vert \psi_{\mathbf{k}}\vert^2}}$ defines the Cooper-pair unoccupied probability, and $v_{\mathbf{k}}=\frac{\psi_{\mathbf{k}}}{\sqrt{1+\vert \psi_{\mathbf{k}}\vert^2}}$ describes the occupied probability. Thus the inner-pair wave function can be defined by $\frac{v_{\mathbf{k}}}{u_{\mathbf{k}}}$, whose absolute value can be obtained by the factors $\vert u_{\mathbf{k}}v_{\mathbf{k}}\vert^2$, $\vert v_{\mathbf{k}}\vert^4$ and $\vert u_{\mathbf{k}}\vert^4$ in the three contributions to $\Gamma^{(2)}$. 

At low temperature, $\Gamma^{(2)}_3$ has little contribution to the cARPES due to the opening of the superconducting gap. In this case, the coincidence probability with a finite center-of-mass energy transfer is defined by $\Gamma^{(2)}_2$, which has a simple relation to the single-particle ARPES counting probabilities:
\begin{equation}
\Gamma^{(2)}_2 = \Gamma^{(1)}\left( g_1 \right)  \Gamma^{(1)}\left( g_2 \right) , \label{eqn3.2.10}  
\end{equation}
where $\Gamma^{(1)}\left( g_1 \right)$ and $\Gamma^{(1)}\left( g_2 \right)$ are the two independent ARPES counting probabilities defined at low temperature as $\Gamma^{(1)} (g) = \frac{2\pi g^2 \Delta T}{\hbar} \vert v_{\mathbf{k}} \vert^2 \delta ( E^{(1)} + E_{\mathbf{k}} ) n_F (-E_{\mathbf{k}})$. As $\Gamma^{(2)}_2$ comes from the propagation of two Bogoliubov quasiparticles, this relation is consistent with the fact that the Bogoliubov quasiparticles are free in the BCS superconducting state defined by the mean-field Hamiltonian Eq. (\ref{eqn3.2.1}).

\section{ $\text{c}$ARP/IPES, $\text{c}$ARIPES and contour-time ordering formalism }\label{sec4}

In Sec. \ref{sec2} we have proposed a cARPES, which can provide the two-particle Bethe-Salpeter wave function in the particle-particle channel. In this section, we will propose another two experimental techniques, a cARP/IPES and a cARIPES. A cARP/IPES shows the two-particle Bethe-Salpeter wave function in the particle-hole channel and a cARIPES involves the two-particle Bethe-Salpeter wave function in the particle-particle channel with the electronic states mainly above the Fermi energy. We will also give a simple discussion on a contour-time ordering formalism for the coincidence detections.

\subsection{ $\text{c}$ARP/IPES }\label{sec4.1}

\begin{figure}[ht]
\includegraphics[width=0.9\columnwidth]{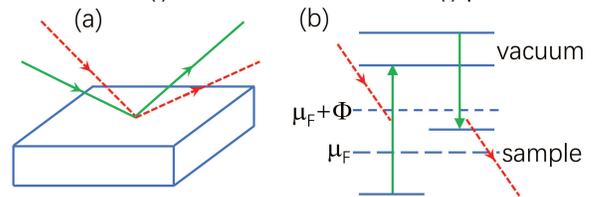} 
\caption{ (Color online) Schematic figures of the cARP/IPES. In (a), the red dashed line with the arrow to the sample represents the incident photon and the green solid line with the arrow to the sample denotes the incident electron. The red dashed line and the green solid line with the arrows outside the sample represent the emitting photon and photoelectron, respectively. (b) The energetics of the cARP/IPES. The symbols are the same as those in Fig. \ref{fig2.3.1}. }
\label{fig4.1.1}
\end{figure}

Fig. \ref{fig4.1.1} shows the schematic diagram and energetics of a cARP/IPES. There are two sources in a cARP/IPES, one for the photon and the other one for the electron. The incident photon can be absorbed by a sample electron which can then be excited into a high-energy state and escape into vacuum to be a photoelectron. The incident electron can transit into a low-energy state of the sample electrons with an additional photon emitting outside into vacuum. The two relevant physical processes can be described by the following electron-photon interaction vertices: 
\begin{eqnarray}
V^{(3)}_1 &=&  g\left(\mathbf{k}_1;\mathbf{q}_1\lambda_1\right) d^{\dag}_{\mathbf{k}_1+\mathbf{q}_1 \sigma_1} c_{\mathbf{k}_1\sigma_1} a_{\mathbf{q}_1\lambda_1} , \notag \\
V^{(3)}_2 &=&  g\left(\mathbf{k}_2;\mathbf{q}_2\lambda_2\right) c^{\dag}_{\mathbf{k}_2\sigma_2} a^{\dag}_{\mathbf{q}_2\lambda_2} d_{\mathbf{k}_2+\mathbf{q}_2 \sigma_2}  , \notag 
\end{eqnarray}
where $V^{(3)}_1$ describes the photoelectric process of photon absorption and photoelectron emission, and $V^{(3)}_2$ describes the transition of the incident electron into a sample electron and the corresponding photon emission. Here we have made a similar approximation to the three-step model with the sudden approximation\citep{ShenRMP2003,BerglundPR1964,FeibelmanPRB1974}  for the photoelectric effect described by $V^{(3)}_1$. For the physical process of $V^{(3)}_2$, we have also made a similar approximation, where the incident electron tunnels into the sample surface and then moves into the sample bulk without interaction with the sample material. 

In a cARP/IPES, the emitting photoelectron and photon are detected by a coincidence detector which records a finite counting when both the emitting photoelectron and photon are detected simultaneously. The coincidence detection probability is defined by 
\begin{equation}
\Gamma^{(3)} = \frac{1}{Z}\sum_{\alpha\beta} e^{-\beta E_\alpha}  \arrowvert \langle \Phi_\beta \arrowvert S^{(3)}(+\infty,-\infty) \arrowvert  \Phi_\alpha \rangle  \arrowvert^{2} , \label{4.1.1}
\end{equation} 
where  
$\arrowvert  \Phi_\alpha \rangle = \arrowvert  \Psi^{s}_\alpha \rangle \otimes  \arrowvert  1_{\mathbf{q}_1\lambda_1}\rangle _p \otimes \arrowvert  1_{\mathbf{k}_2+\mathbf{q}_2 \sigma_2} \rangle_d$ and  $\arrowvert  \Phi_\beta \rangle = \arrowvert  \Psi^{s}_\beta \rangle \otimes  \arrowvert  1_{\mathbf{q}_2\lambda_2}\rangle_p  \otimes \arrowvert  1_{\mathbf{k}_1+\mathbf{q}_1 \sigma_1} \rangle_d$. The relevant $S$ matrix is defined as
\begin{widetext}
\begin{equation}
S^{(3)}(+\infty,-\infty) = \left(-\frac{i}{\hbar}\right)^2 \iint^{+\infty}_{-\infty} T_{t} \lbrack V^{(3)}_{2,I}(t_2) V^{(3)}_{1,I}(t_1) \rbrack F(t_2) F(t_1)  dt_2 dt_1 , \label{eqn4.1.2}
\end{equation}
where $V^{(3)}_{i,I}(t) = e^{i H_0 t/\hbar} V^{(3)}_i e^{-i H_0 t/\hbar}$.

Following a similar procedure to study the cARPES, we introduce a Bethe-Salpeter wave function defined in the particle-hole channel:  
\begin{equation}
\phi^{(3)}_{\alpha\beta}\left( \mathbf{k}_1\sigma_1 t_1; \mathbf{k}_2\sigma_2 t_2 \right) = \langle \Psi^{s}_{\beta} \arrowvert T_t c^{\dag}_{\mathbf{k}_2 \sigma_2} \left(t_2\right) c_{\mathbf{k}_1 \sigma_1} \left(t_1\right) \arrowvert \Psi^{s}_{\alpha} \rangle . \label{eqn4.1.3}
\end{equation}
It describes the physics of the sample electrons when one particle and one hole are created in time ordering, thus it describes the particle-hole pair dynamical physics of the sample electrons.
For the case with a finite but large $\Delta T$, the coincidence probability of the cARP/IPES can be given by 
\begin{equation}
\Gamma^{(3)} = \frac{\left( g_1 g_2 \right)^2 }{(2\pi \hbar)^4} \frac{1}{Z}\sum_{\alpha\beta} e^{-\beta E_{\alpha}} \bigg\vert \iint^{+\infty}_{-\infty} d\Omega	 d\omega \phi^{(3)}_{\alpha\beta}\left(\mathbf{k}_1\sigma_1, \mathbf{k}_2\sigma_2; \Omega, \omega \right) Y(\Omega,\omega) \bigg\vert^2 , \label{eqn4.1.4}
\end{equation}
where $\phi^{(3)}_{\alpha\beta}\left(\mathbf{k}_1\sigma_1, \mathbf{k}_2\sigma_2; \Omega, \omega \right)$ is the frequency Fourier transformation of the Bethe-Salpeter wave function with the center-of-mass and the inner-pair relative time variables. $Y(\Omega,\omega)$ is similarly defined in Eq. (\ref{eqn2.3.13}) with the transfer energies $E^{(2)}$ and $\mathcal{E}^{(2)}$ substituted by $E^{(3)}$ and $\mathcal{E}^{(3)}$ which are defined as 
\begin{equation}
E^{(3)} = E_1^{(3)} + E_2^{(3)} , \mathcal{E}^{(3)} = \frac{1}{2} ( E_2^{(3)} - E_1^{(3)} ) . \label{eqn4.1.5}
\end{equation}  
Here the transfer energies $ E^{(3)}_1$ and $ E^{(3)}_2$ are given by
\begin{eqnarray}
&& E^{(3)}_1 = \varepsilon^{(d)}_{\mathbf{k}_1+\mathbf{q}_1 \sigma_1} + \Phi - \hbar \omega_{\mathbf{q}_1} ,  
E^{(3)}_2 = \hbar \omega_{\mathbf{q}_2} + \Phi - \varepsilon^{(d)}_{\mathbf{k}_2+\mathbf{q}_2 \sigma_2} . \label{eqn4.1.6} 
\end{eqnarray}

The frequency Bethe-Salpeter wave function for the cARP/IPES also has a general form:  
\begin{equation}
\phi^{(3)}_{\alpha\beta}\left(\mathbf{k}_1\sigma_1, \mathbf{k}_2\sigma_2; \Omega, \omega \right)  = 2\pi \delta \left[\Omega + \left( E_{\beta} - E_{\alpha}\right)/\hbar \right] \phi^{(3)}_{\alpha\beta}\left(\mathbf{k}_1\sigma_1, \mathbf{k}_2\sigma_2; \omega \right) , \label{eqn4.1.7} 
\end{equation} 
where $\phi^{(3)}_{\alpha\beta}\left(\mathbf{k}_1\sigma_1, \mathbf{k}_2\sigma_2; \omega \right)$ follows
\begin{equation}
\phi^{(3)}_{\alpha\beta}\left(\mathbf{k}_1\sigma_1, \mathbf{k}_2\sigma_2; \omega \right) = \sum_{\gamma} \left[ \frac{ i \langle\Psi^{s}_{\beta} \vert c^{\dag}_{\mathbf{k}_2 \sigma_2} \vert \Psi^s_{\gamma} \rangle \langle \Psi^s_{\gamma} \vert  c_{\mathbf{k}_1 \sigma_1} \vert \Psi^s_{\alpha} \rangle} {\omega + i\delta^+ + (E_{\alpha} + E_{\beta} - 2 E_{\gamma} )/2\hbar} + \frac{ i \langle\Psi^{s}_{\beta} \vert c_{\mathbf{k}_1 \sigma_1} \vert \Psi^s_{\gamma} \rangle \langle \Psi^s_{\gamma} \vert  c^{\dag}_{\mathbf{k}_2 \sigma_2} \vert \Psi^s_{\alpha} \rangle} {\omega - i\delta^+ - (E_{\alpha} + E_{\beta} - 2 E_{\gamma} )/2\hbar} \right] . \label{eqn4.1.8}
\end{equation}
Now the coincidence probability of the cARP/IPES can be calculated from the following expression:
\begin{equation}
\Gamma^{(3)} = \frac{\left( g_1 g_2 \right)^2 \Delta T}{2\pi \hbar^3} \frac{1}{Z}\sum_{\alpha\beta} e^{-\beta E_{\alpha}} \delta ( E^{(3)} + E_{\beta} - E_{\alpha} ) I^{(3)}_{\alpha\beta}\left(\mathbf{k}_1\sigma_1, \mathbf{k}_2\sigma_2  \right) , \label{eqn4.1.9}
\end{equation}
where $I^{(3)}_{\alpha\beta}\left(\mathbf{k}_1\sigma_1, \mathbf{k}_2\sigma_2  \right)$ is defined by 
\begin{equation}
I^{(3)}_{\alpha\beta}\left(\mathbf{k}_1\sigma_1, \mathbf{k}_2\sigma_2  \right) = \bigg\vert \int^{+\infty}_{-\infty} d\omega \phi^{(3)}_{\alpha\beta}\left(\mathbf{k}_1\sigma_1, \mathbf{k}_2\sigma_2; \omega \right) \frac{2 \sin  \lbrack ( \omega - \mathcal{E}^{(3)}/\hbar )\Delta T/2 \rbrack }{\omega - \mathcal{E}^{(3)}/\hbar } \bigg\vert^2 . \label{eqn4.1.10}
\end{equation}
\end{widetext}

In the limit $\Delta T \rightarrow +\infty$, the coincidence probability of the cARP/IPES has a simple form:
\begin{equation}
\Gamma^{(3)} = \frac{\left( g_1 g_2 \right)^2 }{ \hbar^4} \frac{1}{Z} \sum_{\alpha\beta} e^{-\beta E_{\alpha}} \big\vert \phi^{(3)}_{\alpha\beta}\left(\mathbf{k}_1\sigma_1, \mathbf{k}_2\sigma_2; \Omega_c, \omega_r \right) \big\vert^2 , \label{eqn4.1.11}
\end{equation}
where the frequencies $\Omega_c$ and $\omega_r$ are set by the transfer energies as
\begin{equation}
\Omega_c = E^{(3)}/\hbar, \omega_r = \mathcal{E}^{(3)}/\hbar . \label{eqn4.1.12}
\end{equation}
Obviously, the coincidence probability of the cARP/IPES provides the information on the frequency Bethe-Salpeter wave function in the particle-hole channel. Similarly to the cARPES, the Bethe-Salpeter wave function in the cARP/IPES involves the following particle-hole pair physics of the sample electrons, (1) the pair center-of-mass dynamical physics described by $\delta [\Omega + (E_{\beta}-E_{\alpha})/\hbar]$, and (2) the inner-pair relative dynamical physics which has the resonancelike peak structures at $\omega = \pm (E_{\alpha} + E_{\beta} - 2 E_{\gamma} )/2\hbar$ with the weights defined by $\langle\Psi^{s}_{\beta} \vert c^{\dag}_{\mathbf{k}_2 \sigma_2} \vert \Psi^s_{\gamma} \rangle \langle \Psi^s_{\gamma} \vert  c_{\mathbf{k}_1 \sigma_1} \vert \Psi^s_{\alpha} \rangle$ and $\langle\Psi^{s}_{\beta} \vert c_{\mathbf{k}_1 \sigma_1} \vert \Psi^s_{\gamma} \rangle \langle \Psi^s_{\gamma} \vert  c^{\dag}_{\mathbf{k}_2 \sigma_2} \vert \Psi^s_{\alpha} \rangle$. Therefore, the cARP/IPES is one momentum and energy resolved technique to study the two-particle correlations in the particle-hole channel with both the center-of-mass and inner-pair relative dynamics. As the itinerant magnetism in the metallic ferromagnet/antiferromagnet can be regarded as the physics of the particle-hole pairs in the spin channel and the metallic nematic state\citep{Fradkin,SuLi2015,SuLi2017} is dominated by the particle-hole pairs in the charge channel, the cARP/IPES will play vital roles in the study of the particle-hole pair correlations in these metallic ferromagnet/antiferromagnet and nematic state.

\subsection{ $\text{c}$ARIPES }\label{sec4.2}

\begin{figure}[ht]
\includegraphics[width=0.9\columnwidth]{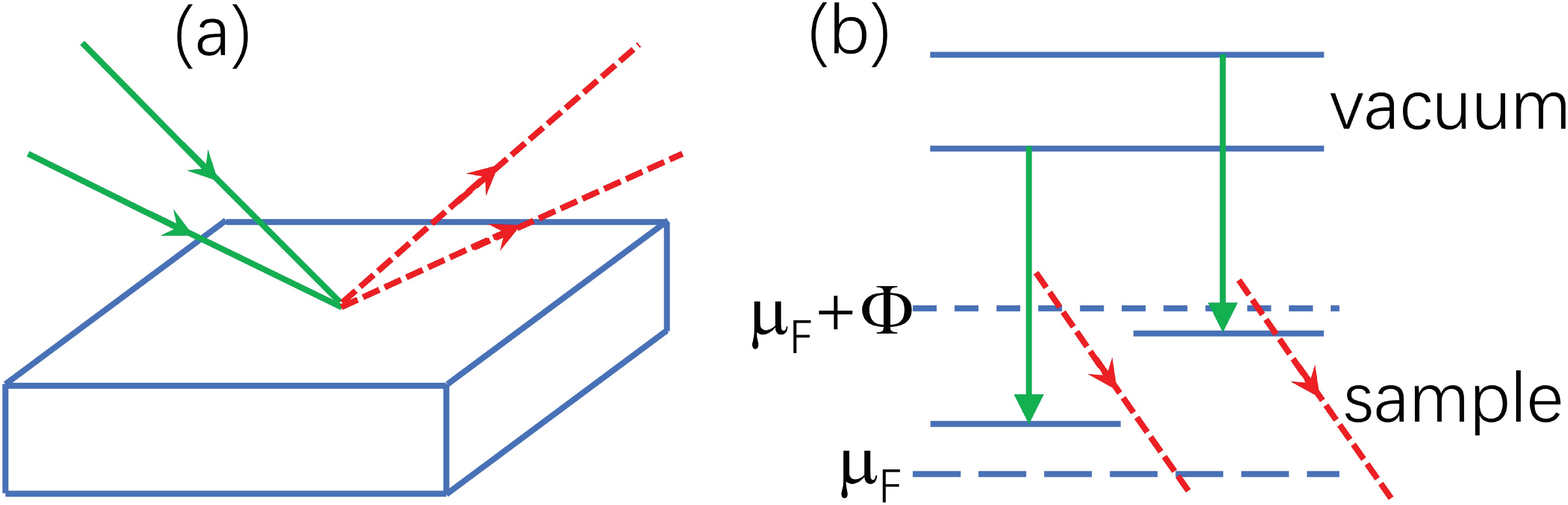} 
\caption{ (Color online) Schematic figures of the cARIPES. In (a), the two green solid lines represent two incident electrons and the two red dashed lines denote two emitting photons. (b) The relevant energetics with the symbols defined the same as those in Fig. \ref{fig2.3.1}. }
\label{fig4.2.1}
\end{figure}

In Fig. \ref{fig4.2.1} we propose another experimental coincidence technique, a cARIPES. In this technique, two electrons are incident on the sample material and transit into the low-energy states of the sample electrons with two additional photons emitting into vacuum. These two emitting photons are then detected in coincidence by a coincidence detector. 

Following the similar approximate three-step model with the sudden approximation\citep{ShenRMP2003,BerglundPR1964,FeibelmanPRB1974} introduced for the ARPES, the cARPES, and the cARP/IPES, the electron-photon interaction vertices for the two respective physical processes in the cARIPES are defined by    
\begin{eqnarray}
V^{(4)}_1 &=&  g\left(\mathbf{k}_1;\mathbf{q}_1\lambda_1\right) c^{\dag}_{\mathbf{k}_1\sigma_1} a^{\dag}_{\mathbf{q}_1\lambda_1}  d_{\mathbf{k}_1+\mathbf{q}_1 \sigma_1} , \notag \\
V^{(4)}_2 &=&  g\left(\mathbf{k}_2;\mathbf{q}_2\lambda_2\right) c^{\dag}_{\mathbf{k}_2 \sigma_2} a^{\dag}_{\mathbf{q}_2\lambda_2} d_{\mathbf{k}_2+\mathbf{q}_2 \sigma_2} . \notag 
\end{eqnarray}

The coincidence detection probability of the two emitting photons in the cARIPES is defined by 
\begin{equation}
\Gamma^{(4)} = \frac{1}{Z}\sum_{\alpha\beta} e^{-\beta E_\alpha}  \arrowvert \langle \Phi_\beta \arrowvert S^{(4)}(+\infty,-\infty) \arrowvert  \Phi_\alpha \rangle  \arrowvert^{2} , \label{4.2.1}
\end{equation} 
where  
$\arrowvert  \Phi_\alpha \rangle = \arrowvert  \Psi^{s}_\alpha \rangle \otimes  \arrowvert  0 \rangle_p  \otimes \arrowvert  1_{\mathbf{k}_1+\mathbf{q}_1 \sigma_1} 1_{\mathbf{k}_2+\mathbf{q}_1 \sigma_2} \rangle_d$ and  $\arrowvert  \Phi_\beta \rangle = \arrowvert  \Psi^{s}_\beta \rangle \otimes  \arrowvert 1_{\mathbf{q}_1\lambda_1} 1_{\mathbf{q}_2\lambda_2}\rangle_p  \otimes \arrowvert 0 \rangle_d$. The $S$ matrix is given by
\begin{widetext}
\begin{equation}
S^{(4)}(+\infty,-\infty) = \left(-\frac{i}{\hbar}\right)^2 \iint^{+\infty}_{-\infty} T_{t} \lbrack V^{(4)}_{2,I}(t_2) V^{(4)}_{1,I}(t_1) \rbrack F(t_2) F(t_1)  dt_2 dt_1 , \label{eqn4.2.2}
\end{equation}
where $V^{(4)}_{i,I}(t) = e^{i H_0 t/\hbar} V^{(4)}_i e^{-i H_0 t/\hbar}$.

With a similar study to the cARPES and the cARP/IPES, we introduce a Bethe-Salpeter wave function defined in the particle-particle channel:  
\begin{equation}
\phi^{(4)}_{\alpha\beta}\left( \mathbf{k}_1\sigma_1 t_1; \mathbf{k}_2\sigma_2 t_2 \right) = \langle \Psi^{s}_{\beta} \arrowvert T_t c^{\dag}_{\mathbf{k}_2 \sigma_2} \left(t_2\right) c^{\dag}_{\mathbf{k}_1 \sigma_1} \left(t_1\right) \arrowvert \Psi^{s}_{\alpha} \rangle . \label{eqn4.2.3}
\end{equation}
It describes the physics of the sample electrons when two particles are created in time ordering. Therefore, it describes the particle-particle pair dynamical physics of the sample electrons. 
The corresponding frequency Bethe-Salpeter wave function is denoted by $\phi^{(4)}_{\alpha\beta}\left(\mathbf{k}_1\sigma_1, \mathbf{k}_2\sigma_2; \Omega, \omega \right)$ with the center-of-mass and the inner-pair relative frequency variables. It follows  
\begin{equation}
\phi^{(4)}_{\alpha\beta}\left(\mathbf{k}_1\sigma_1, \mathbf{k}_2\sigma_2; \Omega, \omega \right)  = 2\pi \delta \left[\Omega + \left( E_{\beta} - E_{\alpha}\right)/\hbar \right] \phi^{(4)}_{\alpha\beta}\left(\mathbf{k}_1\sigma_1, \mathbf{k}_2\sigma_2; \omega \right) , \label{eqn4.2.4} 
\end{equation} 
where $\phi^{(4)}_{\alpha\beta}\left(\mathbf{k}_1\sigma_1, \mathbf{k}_2\sigma_2; \omega \right)$ has a general form:  
\begin{equation}
\phi^{(4)}_{\alpha\beta}\left(\mathbf{k}_1\sigma_1, \mathbf{k}_2\sigma_2; \omega \right) = \sum_{\gamma} \left[ \frac{ i \langle\Psi^{s}_{\beta} \vert c^{\dag}_{\mathbf{k}_2 \sigma_2} \vert \Psi^s_{\gamma} \rangle \langle \Psi^s_{\gamma} \vert  c^{\dag}_{\mathbf{k}_1 \sigma_1} \vert \Psi^s_{\alpha} \rangle} {\omega + i\delta^+ + (E_{\alpha} + E_{\beta} - 2 E_{\gamma} )/2\hbar} + \frac{ i \langle\Psi^{s}_{\beta} \vert c^{\dag}_{\mathbf{k}_1 \sigma_1} \vert \Psi^s_{\gamma} \rangle \langle \Psi^s_{\gamma} \vert  c^{\dag}_{\mathbf{k}_2 \sigma_2} \vert \Psi^s_{\alpha} \rangle} {\omega - i\delta^+ - (E_{\alpha} + E_{\beta} - 2 E_{\gamma} )/2\hbar} \right] . \label{eqn4.2.5}
\end{equation}

With a finite but large $\Delta T$, the coincidence probability of the cARIPES is shown to follow
\begin{equation}
\Gamma^{(4)} = \frac{\left( g_1 g_2 \right)^2 \Delta T}{2\pi \hbar^3} \frac{1}{Z}\sum_{\alpha\beta} e^{-\beta E_{\alpha}} \delta ( E^{(4)} + E_{\beta} - E_{\alpha} ) I^{(4)}_{\alpha\beta}\left(\mathbf{k}_1\sigma_1, \mathbf{k}_2\sigma_2  \right) , \label{eqn4.2.6}
\end{equation}
where $I^{(4)}_{\alpha\beta}\left(\mathbf{k}_1\sigma_1, \mathbf{k}_2\sigma_2  \right)$ is given by 
\begin{equation}
I^{(4)}_{\alpha\beta}\left(\mathbf{k}_1\sigma_1, \mathbf{k}_2\sigma_2  \right) = \bigg\vert \int^{+\infty}_{-\infty} d\omega \phi^{(4)}_{\alpha\beta}\left(\mathbf{k}_1\sigma_1, \mathbf{k}_2\sigma_2; \omega \right) \frac{2 \sin  \lbrack ( \omega - \mathcal{E}^{(4)}/\hbar )\Delta T/2 \rbrack }{\omega - \mathcal{E}^{(4)}/\hbar } \bigg\vert^2 . \label{eqn4.2.7}
\end{equation}
Here the energies $E^{(4)}$ and $\mathcal{E}^{(4)}$ are defined by
\begin{equation}
E^{(4)} = E_1^{(4)} + E_2^{(4)} , \mathcal{E}^{(4)} = \frac{1}{2} ( E_2^{(4)} - E_1^{(4)} ) , \label{eqn4.2.8}
\end{equation}  
with $ E^{(4)}_1$ and $ E^{(4)}_2$ given by
\begin{equation}
E^{(4)}_1 = \hbar \omega_{\mathbf{q}_1} + \Phi - \varepsilon^{(d)}_{\mathbf{k}_1+\mathbf{q}_1 \sigma_1} ,  
E^{(4)}_2 = \hbar \omega_{\mathbf{q}_2} + \Phi - \varepsilon^{(d)}_{\mathbf{k}_2+\mathbf{q}_2 \sigma_2} . \label{eqn4.2.9}
\end{equation}
\end{widetext}

Let us consider the case with the limit $\Delta T \rightarrow +\infty$. In this case, the coincidence probability of the cARIPES has a simple behavior as 
\begin{equation}
\Gamma^{(4)} = \frac{\left( g_1 g_2 \right)^2 }{ \hbar^4} \frac{1}{Z} \sum_{\alpha\beta} e^{-\beta E_{\alpha}} \big\vert \phi^{(4)}_{\alpha\beta}\left(\mathbf{k}_1\sigma_1, \mathbf{k}_2\sigma_2; \Omega_c, \omega_r \right) \big\vert^2 , \label{eqn4.2.10}
\end{equation}
where the transfer energies define the frequencies as
\begin{equation}
\Omega_c = E^{(4)}/\hbar, \omega_r = \mathcal{E}^{(4)}/\hbar . \label{eqn4.2.11}
\end{equation}
It is obviously that the coincidence probability of the cARIPES shows the information on the frequency Bethe-Salpeter wave function in the particle-particle channel. In contrast to the coincidence probability of the cARPES, the relevant particle-particle channel in the cARIPES involves mainly the electronic states above the Fermi energy. This can be easily shown from the definition of the Bethe-Salpeter wave function, Eq. (\ref{eqn4.2.3}). Therefore, the cARIPES can provide the particle-particle correlations with the particles mainly in the states above the Fermi energy. Similar to the cARPES and the cARP/IPES, the particle-particle correlations in the cARIPES involve the pair center-of-mass dynamical physics defined by $\delta [\Omega + (E_{\beta}-E_{\alpha})/\hbar]$, and the inner-pair dynamical physics with the resonancelike peak structures at $\omega = \pm (E_{\alpha} + E_{\beta} - 2 E_{\gamma} )/2\hbar$ which have weights defined by $\langle\Psi^{s}_{\beta} \vert c^{\dag}_{\mathbf{k}_2 \sigma_2} \vert \Psi^s_{\gamma} \rangle \langle \Psi^s_{\gamma} \vert  c^{\dag}_{\mathbf{k}_1 \sigma_1} \vert \Psi^s_{\alpha} \rangle$ and $\langle\Psi^{s}_{\beta} \vert c^{\dag}_{\mathbf{k}_1 \sigma_1} \vert \Psi^s_{\gamma} \rangle \langle \Psi^s_{\gamma} \vert  c^{\dag}_{\mathbf{k}_2 \sigma_2} \vert \Psi^s_{\alpha} \rangle$. If the spin states of the incident electrons can be defined definitely, the cARIPES will be one momentum-energy-spin resolved technique to study the particle-particle correlations of the sample electrons with the electron energies mainly above the Fermi level.

\subsection{ Contour-time ordering formalism}\label{sec4.3}

In the above three coincidence techniques to detect the two-particle correlations, the coincidence probabilities involve the Bethe-Salpeter wave functions which show the momentum and energy dependent physics of the sample electrons in the particle-particle or particle-hole channel. In this section we will show that the coincidence probability can be reexpressed into a contour-time ordering formalism. 

\begin{figure}[ht]
\includegraphics[width=0.6\columnwidth]{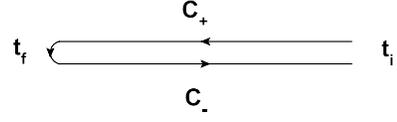}
\caption{ Two-branch contour $C$ for the time ordering operator $T_c$.\citep{SuRaman2016} $t_i$ and $t_f$ are the respective initial and final times. The whole time contour $C$ involves an upper time branch $C_{+}$ and a lower time branch $C_{-}$. If $t_i \rightarrow -\infty, t_f \rightarrow +\infty$, the contour $C$ is the so-called Schwinger-Keldysh contour.\citep{Rammer} }
\label{fig4.3.1}
\end{figure}

Consider the coincidence probability of the cARPES, Eq. (\ref{2.3.1}). This coincidence probability can be reexpressed as following:
\begin{widetext}
\begin{eqnarray}
\Gamma^{(2)} &=& \frac{1}{Z}\sum_{\alpha\beta} e^{-\beta E_{\alpha}} \langle \Phi_{\alpha} \arrowvert S^{(2)}(-\infty,+\infty) \arrowvert  \Phi_{\beta} \rangle   \langle \Phi_{\beta} \arrowvert S^{(2)}(+\infty,-\infty) \arrowvert  \Phi_{\alpha} \rangle   \notag \\
&=&  \frac{1}{Z}\sum_{\alpha} e^{-\beta E_\alpha} 
\left( -\frac{i}{\hbar}\right)^4 \iint_{+\infty}^{-\infty} d t_2^{\prime} d t_1^{\prime} \iint_{-\infty}^{+\infty} d t_2 d t_1 \langle\Phi_\alpha \arrowvert [T_{t^{\prime}} V^{\dag}_{1,I}\left(t_1^{\prime}\right) V^{\dag}_{2,I}\left(t_2^{\prime}\right)  ] \left[T_{t} V_{2,I}\left(t_2 \right) V_{1,I}\left(t_1\right)  \right] \arrowvert \Phi_\alpha \rangle ,   \notag
\end{eqnarray}
where $T_t$ defines the time ordering along $-\infty \rightarrow +\infty$, and $T_{t^{\prime}}$ defines the anti-time ordering along $+\infty \rightarrow -\infty$. $\Gamma^{(2)}$ can be reexpressed into the form by a contour-time ordering: 
\begin{equation}
\Gamma^{(2)} =   \left(-\frac{i}{\hbar}\right)^4 \int_{\left[t_1 t_2;t_1^{\prime} t_2^{\prime}\right]} d t_2^{\prime} d t_1^{\prime} d t_2 d t_1 \langle [T_{c} V^{\dag}_{1,I}\left(t_1^{\prime}\right) V^{\dag}_{2,I}\left(t_2^{\prime} \right) V_{2,I}\left(t_2 \right) V_{1,I}\left(t_1\right) ] \rangle . \label{eqn4.3.1}  
\end{equation}
\end{widetext}
Here $T_c$ is a contour-time ordering operator. It is defined on the time contour $C=C_{+}\cup C_{-}$, where $t\in C_{+}$ evolves as $-\infty \rightarrow +\infty$ and $t^{\prime}\in C_{-}$ evolves as $+\infty \rightarrow -\infty$ as shown schematically in Fig. \ref{fig4.3.1}. The definition of $T_c$ is given by\citep{Rammer,SuRaman2016} 
\begin{equation}
T_c [A(t_1) B(t_2)] = \left\{
\begin{array} {l l l}
A(t_1) B(t_2) , &  \text{if} & t_1 >_c t_2 , \\
\pm B(t_2) A(t_1) , &  \text{if} & t_1 <_c t_2 ,
\end{array}
\right. \label{eqn4.3.2}
\end{equation}
where $>_c$ and $<_c$ are defined according to the position of the time arguments, latter or earlier in the time contour $C$, and $\pm$ are defined for the bosonic or fermionic operators, respectively. 
In Eq. (\ref{eqn4.3.1}), $\left[t_1 t_2; t_1^{\prime} t_2^{\prime}\right] \equiv t_1,t_2 \in C_{+}$ and $t_1^{\prime},t_2^{\prime} \in C_{-} $, and $\langle A \rangle = \frac{1}{Z}\text{Tr} (e^{-\beta H_0} A)$.

In the particle-particle channel for a Cooper pair with $(\mathbf{k}\uparrow,-\mathbf{k}\downarrow)$, the coincidence probability of the cARPES follows  
\begin{widetext}
\begin{equation}
\Gamma^{(2)} =  \frac{\left( g_1 g_2 \right)^2 }{\hbar^4}  \int_{\left[t_1 t_2;t_1^{\prime} t_2^{\prime}\right]} d t_2^{\prime} d t_1^{\prime} d t_2 d t_1 \langle [T_{c} c^{\dag}_{\mathbf{k}\uparrow}\left(t_1^{\prime}\right) c^{\dag}_{-\mathbf{k}\downarrow}\left(t_2^{\prime} \right) c_{-\mathbf{k}\downarrow}\left(t_2 \right) c_{\mathbf{k}\uparrow}\left(t_1\right) ] \rangle e^{i E^{(2)}_1 ( t_1 - t_1^{\prime} )/\hbar + i E^{(2)}_2 ( t_2 - t_2^{\prime} )/\hbar } , \label{eqn4.3.3} 
\end{equation}
and in the particle-hole channel, the coincidence probability of the cARP/IPES follows
\begin{equation}
\Gamma^{(3)} = \frac{\left( g_1 g_2 \right)^2 }{\hbar^4}   \int_{\left[t_1 t_2; t_1^{\prime} t_2^{\prime}\right]} d t_2^{\prime} d t_1^{\prime} d t_2 d t_1 \langle [T_{c} c^{\dag}_{\mathbf{k}_1\sigma_1}\left(t_1^{\prime}\right) c_{\mathbf{k}_2 \sigma_2}\left(t_2^{\prime} \right) c^{\dag}_{\mathbf{k}_2\sigma_2}\left(t_2 \right) c_{\mathbf{k}_1\sigma_1}\left(t_1\right)] \rangle e^{i E^{(3)}_1 ( t_1 - t_1^{\prime} )/\hbar + i E^{(3)}_2 ( t_2 - t_2^{\prime} )/\hbar } . \label{eqn4.3.4} 
\end{equation}
\end{widetext}

Obviously, the time evolution in the contour-time formalism shows that the time dynamics are deeply involved in the coincidence probabilities of the proposed two-particle coincidence detection techniques. Thus, they can be introduced to study the time-retarded physics, such as the dynamical formation of the Cooper pairs, the time-retarded physics of the itinerant magnetic moments and the nematic particle-hole pairs. Moreover, with the reexpressed contour-time formalism, we can introduce the well-established contour-time perturbation formalism to calculate the coincidence probabilities in the study of the weak- or intermediate-coupling electrons.

\section{Summary}\label{sec5}

In this article we have proposed an experimental coincidence technique, the cARPES, to study the two-particle correlations. In the cARPES, two incident photons are absorbed and two photoelectrons are emitting into vacuum. A coincidence detector records the two photoelectrons in coincidence with the counting probability relevant to a two-particle Bethe-Salpeter wave function in the particle-particle channel. The cARPES spectra of a free Fermi gas and a BCS superconducting state have been studied in detail.

We have also presented another two experimental coincidence techniques, the cARP/IPES and the cARIPES. In the cARP/IPES, an incident photon excites a photoelectron and an incident electron transits into a low-energy state of the sample electrons with an additional photon emitting into vacuum. The emitting photoelectron and photon are detected in coincidence by a coincidence detector with the coincidence probability relevant to a two-particle Bethe-Salpeter wave function in the spin or charge particle-hole channel. There are two incident electrons in the cARIPES which transit into the low-energy states of the sample electrons with two additional photons emitting into vacuum. A coincidence detector detects the two emitting photons in coincidence, and the counting coincidence probability is relevant to a two-particle Bethe-Salpeter wave function in the particle-particle channel with main contribution from the electronic states above the Fermi energy.  

All of the three experimental coincidence techniques can provide directly the information on the frequency Bethe-Salpeter wave functions in the particle-particle or particle-hole channel. Since the frequency Bethe-Salpeter wave functions show the momentum and energy dependent two-particle dynamical physics of the sample electrons, these coincidence techniques can be introduced to study the momentum and energy resolved two-particle correlations with the center-of-mass and inner-pair relative dynamics. If the spin configurations of the photoelectrons or the incident electrons can be detected, these coincidence detection techniques will be momentum-energy-spin resolved in the study of the two-particle correlations in the particle-particle or particle-hole channel. Moreover, the inner-pair time-retarded physics can also be studied by these coincidence detection techniques. 

The three experimental coincidence techniques proposed to detect the two-particle correlations will play important roles in the study of the many-body physics of the strongly correlated electron materials, such as the microscopic pairing mechanism of the Cooper pairs in the unconventional superconductor, the formation of the itinerant magnetic moments in the metallic ferromagnet/antiferromagnet, and the inner-pair physics of the particle-hole pairs in the metallic nematic state.   

{\it Acknowledgement}
We thank X. Chen, D. Z. Cao, B. Zhu and H. G. Luo for invaluable discussions. This work was supported by the National Natural Science Foundation of China (Grant Nos. 11774299 and 11874318) and the Natural Science Foundation of Shandong Province (Grant Nos. ZR2017MA033 and ZR2018MA043).

\appendix 

\section{Calculation of $\Gamma^{(2)}_1$ in superconducting state} \label{seca1}

From Eq. (\ref{eqn2.3.12}) or (\ref{eqn2.3.14}), the contribution of $\phi^{(2)}_{\alpha\beta,1}$ to the coincidence probability of the cARPES for the BCS superconducting state is shown to follow
\begin{equation}
\Gamma^{(2)}_1 = \frac{2\pi\left( g_1 g_2 \right)^2 \Delta T}{ \hbar^3 }\frac{1}{Z} \sum_{\alpha} e^{-\beta E_{\alpha}} \delta  ( E^{(2)} ) \vert u_{\mathbf{k}} v_{\mathbf{k}} \vert^2 I_{\alpha} , \notag 
\end{equation}
where $I_{\alpha}$ is defined by
 \begin{equation}
I_{\alpha} =  \vert c_1 n_{\mathbf{k}\uparrow}^{\alpha} + c_1^{\ast}(1-n_{\mathbf{k}\uparrow}^{\alpha}) + c_2 (1-n_{-\mathbf{k}\downarrow}^{\alpha}) + c_2^{\ast} n_{-\mathbf{k}\downarrow}^{\alpha} \vert^2  \notag 
\end{equation} 
with $c_1$ and $c_2$ given by
\begin{eqnarray}
&&c_1 = \frac{1-e^{ i(E_{\mathbf{k}}+\mathcal{E}^{(2)} + i \delta^+)\Delta T/2\hbar}} { (E_{\mathbf{k}}+\mathcal{E}^{(2)} + i \delta^+)/\hbar}, \notag \\
&& c_2 = \frac{1-e^{ -i(E_{\mathbf{k}}-\mathcal{E}^{(2)} - i \delta^+)\Delta T/2\hbar}} { (E_{\mathbf{k}}-\mathcal{E}^{(2)} - i \delta^+)/\hbar}. \notag  
\end{eqnarray}
Since $n_{\mathbf{k}\sigma}^{\alpha}=0,1$, $\Gamma^{(2)}_1$ can be furtherly obtained as 
\begin{equation}
\Gamma^{(2)}_1 = \frac{2\pi\left( g_1 g_2 \right)^2 \Delta T}{ \hbar^3 } \delta ( E^{(2)} ) \vert u_{\mathbf{k}} v_{\mathbf{k}} \vert^2 C_I , \label{eqnA.1}
\end{equation}
where $C_I$ is defined as
\begin{eqnarray}
C_I &=&  \vert c_1 + c_2^{\ast} \vert^2 [n^2_F(E_{\mathbf{k}}) + n^2_F(-E_{\mathbf{k}})] \notag \\
&+& 2 \vert c_1 + c_2 \vert^2 n_F(E_{\mathbf{k}}) n_F(-E_{\mathbf{k}}) . \label{eqnA.2}
\end{eqnarray}
In the limit with large $\Delta T $, we have the following results that 
\begin{equation}
\vert c_1 + c_2 \vert^2 = \vert c_1 + c_2^{\ast} \vert^2  = C_{12} \label{eqnA.3}
\end{equation}
with 
\begin{equation}
C_{12} = \pi \hbar\Delta T [ \delta (\mathcal{E}^{(2)}+E_{\mathbf{k}}) + \delta (\mathcal{E}^{(2)}-E_{\mathbf{k}}) ] , \label{eqnA.4}
\end{equation}
which can be shown by mathematical plotting as functions of $\mathcal{E}^{(2)}$ and confirmed partially from Eq. (\ref{eqn2.3.6}) in the limit $\Delta T\rightarrow +\infty$. Substituting these results back into Eq. (\ref{eqnA.1}) and (\ref{eqnA.2}), we can obtain $\Gamma^{(2)}_1$ as shown in Eq. (\ref{eqn3.2.9}), where $n_F(E_{\mathbf{k}}) + n_F(-E_{\mathbf{k}})=1$ has been used. It is noted that the temperature-dependent Fermi distribution function $n_F$ does not appear explicitly in $\Gamma^{(2)}_1$.



\end{document}